\documentclass[sigplan,10pt]{acmart}
\AtBeginDocument{%
  }

\setcopyright{acmlicensed}
\copyrightyear{2018}
\acmYear{2018}
\acmDOI{XXXXXXX.XXXXXXX}
\acmConference[Conference acronym 'XX]{Make sure to enter the correct
  conference title from your rights confirmation email}{June 03--05,
  2018}{Woodstock, NY}
\acmISBN{978-1-4503-XXXX-X/2018/06}

\newcommand{\system}{Codetations}
\begin{document}

\title{\system{}: Intelligent, Persistent Notes and UIs for Programs and Other Documents}

\author{Edward Misback}
\email{misback@cs.washington.edu}
\orcid{0009-0003-9474-0826}  
\affiliation{%
  \institution{University of Washington}
  \city{Seattle}
  \country{USA}}
\author{Erik Vank}
\orcid{0009-0008-7844-6025}
\email{erik_vank@brown.edu}
\affiliation{%
  \institution{Brown University}
  \city{Providence}
  \state{Rhode Island}
  \country{USA}
}
\author{Zachary Tatlock}
\email{ztatlock@cs.washington.edu}
\orcid{0000-0002-4731-0124}
\affiliation{%
  \institution{University of Washington}
  \city{Seattle}
  \country{USA}}
\author{Steven L. Tanimoto}
\email{tanimoto@cs.washington.edu}
\orcid{0000-0002-8175-7456}
\affiliation{%
  \institution{University of Washington}
  \city{Seattle}
  \country{USA}}

\renewcommand{\shortauthors}{Trovato et al.}

\begin{abstract}
Software developers maintain extensive mental models of code they produce and its context, often relying on memory to retrieve or reconstruct design decisions, edge cases, and debugging experiences. These missing links and data obstruct both developers and, more recently, large language models (LLMs) working with unfamiliar code. We present Codetations, a system that helps developers contextualize documents with rich notes and tools. Unlike previous approaches, notes in Codetations stay outside the document to prevent code clutter, attaching to spans in the document using a hybrid edit-tracking/LLM-based method. Their content is dynamic, interactive, and synchronized with code changes. A worked example shows that relevant notes with interactively-collected data improve LLM performance during code repair. In our user evaluation, developers praised these properties and saw significant potential in annotation types that we generated with an LLM in just a few minutes.
\end{abstract}



\begin{teaserfigure}
  \setlength{\fboxsep}{0pt}%
   \setlength{\fboxrule}{0.5pt}%
   \fbox{\includegraphics[width=\dimexpr\textwidth-2\fboxsep-2\fboxrule\relax]{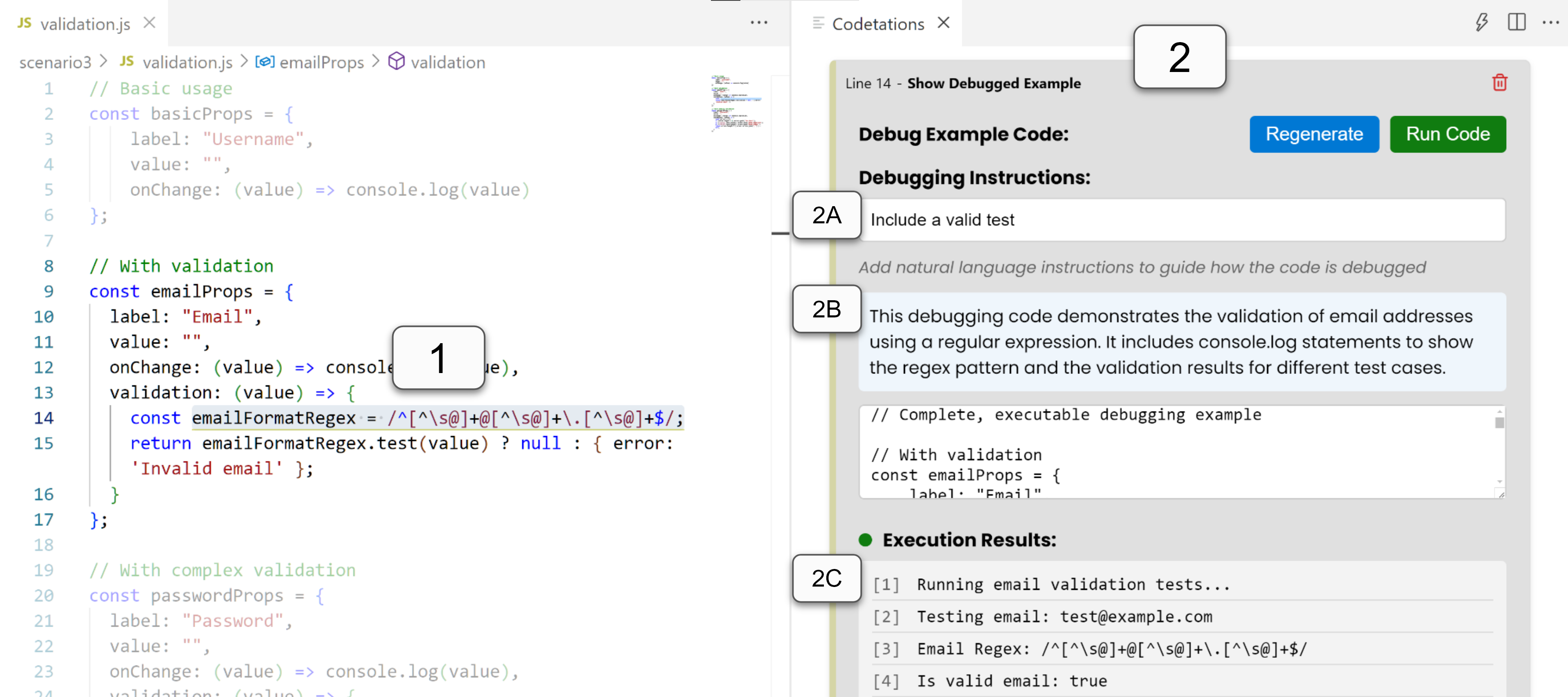}}
  \caption{Codetations lets users attach rich, living annotations to any text file. 
  Here, (1) the behavior of highlighted code is illustrated by (2) a ``Show Debugged Example'' annotation through (2A) user-customizable debugging instructions, (2B) LLM-generated test code, and (2C) live execution results. We added this annotation type to our system in just a few minutes using the prompts shown in Appendix \ref{sec:show-debugged-example-prompt}. Unlike traditional comments and notebooks, Codetations does not modify file content but instead uses a hybrid editor/LLM anchoring system robust to both online and offline edits.}
  \Description{}
  \label{fig:teaser}
\end{teaserfigure}


\received{20 February 2007}
\received[revised]{12 March 2009}
\received[accepted]{5 June 2009}

\maketitle

\section{Introduction}
Software development is inherently contextual. Developers maintain mental models that encompass far more information than exists in source files, including design decision rationales, implementation trade-offs, edge cases, maintenance processes, and debugging workflows; these conceptual models rarely find a permanent home in the codebase itself. With the rise of large language models (LLMs) and other programmatic agents and tools that enhance development workflows, the problem of missing context surfaces not only when code is handed over to new maintainers (or when  programmers forget their own past work) but whenever any agent is invoked without sufficient context, 
contributing to well-known problems like hallucination \cite{martino2023knowledge}.


The problem of missing context cannot be solely attributed to overwhelmed developers who lack time to explain their assumptions. 
Our need-finding work (Section \ref{sec:user-study}) highlights that developers \textit{do} seek ways to maintain development context as a natural part of the programming process: they sketch solutions by hand, maintain personal notes in separate documents, trace through example code executions in REPLs, discuss approaches with teammates, or use other long-form methods for code modeling. Developer attempts to creatively structure this context as code comments can lead to ``messy'' code and hard-to-manipulate representations, such as ASCII art diagrams \cite{hayatpur2024taking}. The reality we find is that \textit{many kinds of context are simply ill-suited to representation within a program's code,} forcing developers to keep it elsewhere and recall when and where it is relevant.




Context is a language-independent property of every codebase, but no widespread, editor-level approach attaches rich context to specific parts of code and documentation. We posit that LLMs may both finally bring this problem to a head and become part of its solution.

We propose that editors assist programmers in linking individual code spans to external documents, executions, or diagrams, including live diagrams. Of course, such systems have been proposed before---but they suffer from a key limitation: how to keep the contextual annotations in sync with code as it changes. Not only does that mean tracking code spans as the code evolves (usually called \textit{anchoring}), but it also means updating the external documents, executions, and diagrams as the code changes (which we call \textit{syncing}). Our insight is that modern LLMs, with their rich semantic understanding of code and documents, allow us to finally surmount this challenge and automatically keep external contextual information in sync with code as the code changes.

Guided by early need-finding and prototype work, we structure our investigation around three questions:
\begin{enumerate}
  \item Do semantically anchored, richly interactive annotations tangibly help programmers---and which parts of the concept matter most?
  \item What contextual information do today’s programmers need to capture and retrieve, and how do they expect those needs to grow?
  \item How much does explicit context improve an LLM's reasoning about code repair issues that it could, in principle, already solve?
\end{enumerate}

As a design probe, we present Codetations, a VSCode extension that hosts interactive annotations and keeps them attached and synced for any code repository. We solve the anchoring problem by tracking code edits in real-time through the editor and employing LLMs to resolve offline updates. We solve the semantic synchronization issue by offering annotation developers an easy-to-use API that lets annotations access host system resources, including, for our implementation, the document text, other annotation data, and editor APIs, including an LLM API.

Our proposed \system{} system allows a programmer to work with a variety of custom UIs for recording, manipulating, and displaying context during a programming workflow; we call these custom UIs \textit{annotation types}. The Codetations API is designed to make new annotation types easy to build: we found that a single request to an LLM with examples of the API's usage was sufficient to immediately add powerful new tools to the system, including the Show Debugged Example annotation (Figure \ref{fig:teaser}). 

In an interview-based user study (N=9) we conducted with experienced programmers to investigate their needs with respect to maintaining code context and responses to a working prototype system (Section \ref{sec:user-study}), participants expressed enthusiasm about the system's features, with all study participants spontaneously suggesting or agreeing that document external annotations would permit greater storage of context with their code and imagining unique ways they could use \system{}-style notes in their own work.

Finally, we found that annotation data created by tools like ours was helpful to an LLM that repairs bugs in code (Section \ref{sec:machine-eval}).


In sum, this paper's contributions include:
\begin{itemize}
    \item \textbf{\system{}}, a system for keeping dynamic, interactive annotations on semantic entities as documents evolve (Section \ref{sec:implementation}).
    \item Results of a \textbf{qualitative user study} exploring developer needs with respect to maintaining code context and developer responses to a working prototype (Section \ref{sec:user-study}).
    \item \textbf{Recommendations} for new annotation types, including examples of rapid annotation generation with LLMs, and specific \textbf{design takeaways} for researchers conducting future work in this area (Section \ref{sec:applications}).
    \item Results of a \textbf{worked example} of LLM performance for debugging tasks with and without attached contextual information (Section \ref{sec:machine-eval}).
\end{itemize}


\section{Related Work}
\label{sec:related-work}

\subsection{Annotations}
Considerable prior work addresses document annotation. Agosti et al. provide a historical study \cite{agosti2007historical} and formal model \cite{agosti2007formal} that clarify the design space for digital annotation management systems.
Brush et al. discuss implementation of robust annotation systems for documents, focusing on  problems like annotation orphaning (the loss of an annotation's position after the anchor text is removed) that require user interactions to resolve \cite{brush2001robust}. We build upon this foundation.

Several efforts attempt to solve the problem of anchoring annotations to positions in code. In 2002, Collard described an XML representation of all code \cite{collard2002supporting} meant for general adoption;
however, it is infeasible today to implement this editor-managed solution globally (but perhaps not locally) for modern programming environments and unstructured documents. 
Reiss's text comparison-based method for finding source code locations after a change \cite{reiss2008tracking} offers the best traditional solution. Horvath et al.'s Catseye system uses this method to attach rich comments to source code \cite{horvath2022catseye}, and their Sodalite system uses it to let users reference source code in documentation \cite{horvath2023sodalite}. However, Horvath et al. note that their method occasionally requires users to manually reattach annotations; in particular, it lacks understanding of what text actually means and cannot handle re-attachments that must account for document  meaning. Related clone detection methods \cite{duala2010clone} suffer from the same problem.

Further, Horvath et al.'s Meta-Manager introduces an automated process for tracking the provenance of code edits that could further enhance annotation systems by attaching edit history and developer interactions with external tools directly to relevant sections of the codebase~\cite{horvath2024meta}. Meta-Manager follows version history of code snippets by parsing the AST syntactically using the method of Wittenhagen et al. \cite{wittenhagen2016chronicler} and is currently restricted to tracking user edits on TypeScript files in the VSCode editor.

Misback et al.~\cite{Misback2024MagicMM} recently described a method for using an LLM to update annotation positions. An LLM's understanding of document semantics allows it to outperform Reiss's method with a simple prompt, matching human expectations for an anchoring system. Our proposed system uses this method despite some weaknesses in its performance when using current LLMs (see Section \ref{sec:implementation} and Appendix \ref{sec:user-study-lm-caveats}).

More recently, popular AI-powered code editors have emerged that address contextual information in different ways. Notably, the Cursor editor \cite{cursorrules} includes ``rule'' annotations, stored in their own subdirectory of a codebase, and an implementation of the Model Context Protocol (MCP) to provide standard ways for developers to add and connect AI assistants with context. Our Codetations system takes a complementary approach by allowing developers to explicitly attach persistent, rich annotations directly to code spans.

\subsection{Annotations as IDE}

Notably, Reiss's work on the Field programming environment \cite{reiss1990connecting} considers  the implications of attaching arbitrary tools to parts of a program. Field annotations are maintained on a document by the editor, and system applications can subscribe to updates from those annotations. A pilot study (Section \ref{sec:design-process}) motivated our user study and implementation to focus on a more constrained editor-embedded system with tighter coupling than Reiss imagines in order to help users recognize possible workflows.

\subsection{Rich Tools for Live and Literate Programming}
\system{} applications must consider the problem of embedding \textit{live} and \textit{rich} tools in codebases as program annotations. Tanimoto \cite{tanimoto2013perspective} describes \textit{live programming systems} as those that can be edited as they are running to obtain immediate feedback on the results of a programmer's actions. Horowitz and Heer \cite{horowitz2023live} define \textit{rich systems} as allowing a programmer to edit programs through domain-specific visualizations. A broad range of systems meet these definitions. Horowitz and Heer \cite{horowitz2023engraft} note that live and rich tools and systems face  challenges with inter-operation; in particular, problems arise when trying to inter-operate with the ``tools and environments in the outside [non-live, rich] world.'' Our system provides a new way to attach these tools to typical codebases.

Many prior works have attempted to integrate live and rich tools with standard programming tools. Computational notebooks are perhaps the most successful example; these are a form of \textit{literate programming} \cite{knuth1984literate} wherein code can be attached to detailed explanations and tools for data visualization. Their typical interface, which lists executable cells (occasionally displayed out-of-order to aid in presentation) resembles  annotations that might be shown in the margin of another document. As such, we considered Lau et al.'s description of the design space of computational notebooks \cite{lau2020design} in the design of our own annotation hosting interface. Unlike \system{}, notebooks require a codebase to be written and accessed using special notebook formats and editor interfaces rather than augmenting existing code and editors.

The recent work of Gobert and Beaudouin-Lafon on Lorgnette \cite{gobert2023lorgnette} details a subtype of live and rich tools called \textit{projections} and an implementation of a projection hosting system with customizable tools very similar to the more general tool hosting system presented here. Unlike Codetations, Lorgnette's tools are attached to all instances of a pattern using regex matching and cannot save the state of an individual instance outside of the code. 

See Section \ref{sec:applications} for additional related work.

\section{Our Design Process}\label{sec:design-process}

We now describe our  
initial motivations and questions regarding system design, discuss early prototypes we discarded, and finally present five key design decisions we made and applied to build our prototype system. 

\subsection{Design Motivation and Research Questions}\label{sec:research-questions}

We have noted that software development involves managing many kinds of contextual information. A one-size-fits-all approach to representing this diverse context is inadequate given the variety of possible use cases and stakeholders (including different kinds of programmer-users and programming agents). We considered the following research questions:

\textbf{RQ1.} Is there any contextual utility in semantically attached, automatically contextualized notes with rich and interactive interfaces? How important are the individual parts of this hypothesized solution?

\textbf{RQ2.} What are the needs of programmer-users today when creating and managing context? What contextual needs do programmer-users anticipate having in the future?

\textbf{RQ3.} How important is context for LLM reasoning? To what extent do current LLMs rely on contextual hints to solve problems whose answers they are capable of reasoning over (e.g., can model A always solve a programming problem designed by model A)?

At a high level, RQ1 probes a possible solution for the challenge of context management, and RQ2 and RQ3 focus on understanding needs in order to further evaluate that solution. We reflect on these questions in Section \ref{sec:discussion}.




\subsection{Early Explorations}

We began our investigation by discussing ideas for the system with colleagues but found it difficult to 
conceptualize and explain the system without a tangible demonstration, so we created a rough working prototype (shown in Appendix Figure~\ref{fig:pilot-ui}). This early implementation was fully editor-independent and browser-based. The interface, while functional, had significant usability challenges, but even beyond these, after explaining the system's purpose and operation, pilot users still had difficulty understanding the system's purpose and imagining realistic workflows using it.

This invaluable experience with our initial prototype prompted us to conduct two rounds of user studies with the current prototype, as described in Section \ref{sec:user-study}. We consider that characterizing the shortcomings in both our early and current implementation constitutes perhaps our most valuable contribution for other annotation system designers.

The primary lesson we learned was that programmers have existing mental models for how annotations should look and work, and our initial system did not meet its users' expectations. To elicit meaningful feedback on our research questions, we decided to focus on identifying existing user workflows and mental models and ensuring that our system provided obvious value within those constraints before asking users to suggest additional possibilities and consider future systems.

As we navigated the design space, we encountered key design questions (below) where this lesson helped to constrain our responses, and we explore these choices in the following sections.

\subsection{Five Key Design Questions}\label{sec:design-decisions}

\subsubsection{Should annotations be hosted through the editor interface and belong to the editor conceptually, or should they reside elsewhere?} 

This question addresses our highest-level design choice. Based on the difference in users' immediate understanding of the system and ability to imagine further capabilities after we rewrote it as an editor extension, we unequivocally find that \textit{users} consider annotations to be an editor feature. But should designers? In this case, the preference of users runs counter to the software design principle of separation of concerns, which tries to keep features independent to allow systems more degrees of freedom. Annotation interfaces for read-only code could certainly be hosted in applications that are not editors, e.g., a browser extension could maintain annotations for Github repositories, and if an annotation interface were fully independent, it could service documents across many different user applications, like Reiss's Field system \cite{reiss1990connecting}.

For our implementation, however, we decided that editor integration provides clear benefits: annotations can (1) appear on the code to which they are attached with their (2) content rendered nearby and (3) respond to edit actions with low latency, matching the mental model that annotation users have formed from document editors like Microsoft Word and Google Docs.

\subsubsection{What kind and customization of annotation types should we host in our prototype system to communicate the system's capabilities to a typical programmer? }\label{sec:what-types} Since contextual information is useful in most parts of the programming process, we first discussed the range of possible annotation data types and workflows that would use them. These choices inform the API a system needs to offer for annotations to work, e.g., a file-writing API is needed for annotation types that modify the region where the annotation is attached.
Section \ref{sec:applications} describes the kinds of annotations we created or proposed and those suggested by participants in our user study.

\subsubsection{How should we connect users' incremental or major file edits to updates in the anchor position and to the content of annotations?}


Broadly, there are trade-offs between correct responses to changes and latency. For example, using an LLM to obtain a more semantically correct update of the position or annotation content is currently relatively expensive in terms of both time and computational energy. Further, a  question arises about whether updates from an LLM should wait for user permission to run in case the results would violate user expectations and require intervention.

Of course, the issue of anchoring could be solved entirely by embedding markers directly in code as comments. However, doing so means mixing the annotation layer into the document, which requires all parties using the document to agree on an acceptable level of code clutter and to avoid accidental damage to markers when documents are edited; it also cannot work in environments without comment support, like JSON files, and requires write access to annotated documents. We think these issues rule out this method for use in an annotation system intended for many kinds of codebases and documents.

Our prototype system tries to strike a balance between these options by responding to online edits that occur through normal editor actions with simple automatic positional updates and reserving the slow, semantically correct, and potentially surprising reattachment method for cases where user permission has been received after an offline update (e.g., when a remote user merges a branch).

\subsubsection{To users, what are system-wide capabilities for working with any annotation vs capabilities for a specific kind of annotation? } Users may request many powerful features for  designers to consider. These choices chiefly affect the information that is stored and must be kept accurate for every annotation.
Annotation anchor information (which describes the document part the annotation refers to) and actual annotation data must always be included. Beyond these essentials, other information can be considered, such as globally unique IDs, order or positions within the set of annotations, author information, timestamps, content hashes, or explicit cross-references between annotations.

Our current system supports only an ID; a single annotation type name, which it uses to render the annotation as a React component; and a component-controlled data field for saving state. This approach still lets annotation types host their own complex behaviors (e.g., cross-referencing could be supported by reading other annotations' data), but system-level behaviors remain relatively basic. Multiple users in our study wanted more than this; in particular, they saw value in annotations that could point to multiple code sections (see Section 5), which would require the formal definition to be expanded to permit multiple anchors.

\subsubsection{Where is the user data stored?}\label{sec:data-where} Though this may seem like an implementation question, the location of users' data directly affects and unifies many user-facing concerns about collaboration, sharing, and version control workflows, and we maintain that users should always know where their data is being kept when using a system. 

When annotation data is stored along with code in the repository, teams immediately gain shared context, and annotations naturally follow version control history. However, such sharing also means that annotations become part of the codebase's footprint and could become a source of merge conflicts.
This decision also affects the basic user experience for teams: repository-stored annotations are simple but require commits for other users to see annotation changes, while database solutions offer real-time collaborative features but introduce authentication complexity. Different storage models also enable different annotation features; some kinds of annotation  may need to persist sensitive information separately from public repositories.

Our prototype uses a JSON file in the repository for simplicity, but using a mixed, local-first approach could let users opt in to private and shared storage, better serving diverse user needs.

\section{System Design}\label{sec:implementation}
\system{} is intended to probe in a structured way the design space issues identified by the questions in Section \ref{sec:design-decisions}. The interface is shown in Figure \ref{fig:teaser}. At the highest level, \system{} allows a programmer working in an editor to add persistent, self-updating, interactive annotations to selected text. An annotation view for displaying and modifying annotations is shown to users in the editor adjacent to (or perhaps interleaved with) the file buffer. 

Appendix \ref{sec:basic-annotations} describes basic features any annotation system should have. Here we distinguish novel features \system{} provides, 
examine its architecture, and provide details about the system version we implemented for our user study.

\subsection{Novel \system{} Feature Set}

\subsubsection{Document-external annotations}\label{sec:document-external} No changes are made to the text of the file when an annotation is added. Instead, the specified anchor points for the annotation are noted in a separate annotation data file.

\subsubsection{Intuitive annotation movement through semantic anchoring}\label{sec:location-updates}
Broadly, the system tries to keep annotations attached to the same semantic entity as the text evolves. As the programmer edits the code, the system updates the anchor points using simple heuristics that match the behavior of comments in popular document editors like Microsoft Word. However, for edits not initiated by the programmer, e.g. when the code is updated on disk by a remote collaborator or a different editing program, \system{} notifies the user of any annotations that have become detached and requests permission to update the anchor points. If the user confirms, the anchor points for each annotation are re-attached to semantically similar points in the new document using the best available method (in our implementation, the method from Misback et al.~\cite{Misback2024MagicMM}). This protects the annotation data from the unavoidable real-world case where remote developers are not running \system{}. Additional measures to ensure anchor consistency, such as periodically checking whether anchor points need to be adjusted, can be considered but were not implemented for our prototype.

\subsubsection{Automatic annotation and file content updates through context-aware annotations}\label{sec:updating-content}
The annotation system itself does not include domain knowledge or heuristics to decide whether annotation or file content is up to date; instead, it delegates this responsibility to the annotation implementations. Annotations receive updates about the document's contents and the section of the document they are attached to, and they are provided with an API (that could be extended with a permissions model) for writing to the document, as needed.

\subsubsection{Interactive ``code''-tations}\label{sec:code-tations}
To probe the full space of possible annotation behaviors mentioned in Section \ref{sec:what-types}, even beyond the file access necessary for \ref{sec:updating-content}, \system{} is designed to host dynamic, interactive annotations with programmatic behaviors. It also offers an API through which the programmatic power of a host system is offered to annotations.

\subsection{\system{} System Architecture}

\subsubsection{Annotation definition} 

Each annotation has a \textit{tag} (i.e., an annotation attachment record) stored independently from the document. 
A \system{} annotation record has the following fields:
\begin{enumerate}
    \item A universally unique identifier for the tag
    \item The start position of the annotation in a document version $D$
    \item The end position of the annotation in $D$
    \item A method for obtaining $D$
    \item An \textit{annotation type} to use for rendering
    \item Data for the annotation 
\end{enumerate}
Elements 2 through 4 anchor the annotation to provide the external anchoring from Section \ref{sec:document-external}; any alternative anchoring method should be able to produce these elements.
List element 5, \textit{annotation types}, are an implementation-dependent mechanism (described in Section \ref{sec:annotation-types}) for allowing the system to provide the content updates and programmatic behaviors described in Sections \ref{sec:updating-content} and \ref{sec:code-tations} above.

\subsubsection{Host and rendering processes}
\system{} consists of a \textit{host or editor-related process} and an \textit{annotation rendering process} (that may render in the editor or elsewhere). The host process provides access to the document or buffer, the annotation data, and any APIs or other system or editor functions required by the rendering process. In our implementation, this host process is a NodeJS VSCode extension. The rendering process gives annotation types access to the functions exposed by the host. In our implementation, the rendering process is a VSCode Webview that can render components written with HTML and JavaScript.

\subsection{Implementation Details}

In this section, we describe the main implementation-dependent features of our prototype. For other implementation details, see Appendix \ref{sec:implementation-extra}.

\subsubsection{Annotation types}\label{sec:annotation-types}  
\textit{Annotation types} are defined in our implementation as React function components that receive as arguments (1) the annotation data and (2) an API for getting and setting the document text and annotation data field from the \system{} host. 

\subsubsection{API for annotation types}
Our implementation gives annotations access to the following annotation type API:
\begin{enumerate}
    \item Functions for reading and writing the annotation's data
    \item Functions for reading and writing the document text
    \item A function for calling the VSCode Language Model API\cite{microsoft_vscode_lm_api}
\end{enumerate}
This API could be extended with additional host access, e.g., to permit use of the host's interpreters or compilers for executing annotated code.

We found that providing a description of this API to a language model allowed us to \textbf{generate complex annotation types with a single prompt}, including the Show Debugged Example annotation type (Section \ref{sec:program-data-live-execution}, prompt in Appendix \ref{sec:show-debugged-example-prompt}) and the LM Unit Test (Section \ref{sec:lm-unit-test}). This capability could be offered to end-users to allow them to create new annotation types for a codebase on-the-fly.

\subsection{Annotation data}
We store tag records in a JSON file that contains a simple array of tag records. Tags are fully independent, i.e., they can be updated in parallel, and they will not be affected if other tags become damaged. This setup also allows annotation types to be written as a function of a single tag record (see Section \ref{sec:annotation-types}).

We place this JSON file in a lazily generated tree stored under a hidden ``.\system{}'' directory kept at the root of its git repository. We noticed that study participants immediately understood how to find the raw annotation data and understood that it would be synced with the rest of their repository. As Section \ref{sec:data-where} notes, keeping annotation data in version control involves significant trade-offs.




\section{User Study}\label{sec:user-study}

To evaluate \system{} and understand developers' contextual information needs, we conducted a user study with a small convenience sample of 9 participants from varied backgrounds including academic researchers, industry developers, and students, with programming experience ranging from 5 to 30 years across domains such as web development, scientific computing, compilers, and SQL database programming. Participant backgrounds are shown in Appendix Table \ref{tab:participant-backgrounds}.

\subsection{Study Design}

Sessions were conducted remotely or in-person, lasted 60-90 minutes, and followed a three-part structure, with about 20 minutes allocated for each part: a pre-demo need-finding discussion and survey focused on documentation practices (see Appendix \ref{sec:pre-survey}); a hands-on demonstration of \system{}, described below; and a post-demo discussion and survey to gather immediate reactions to our system and perspectives on its future (see Appendix \ref{sec:post-survey}). We iterated on the demo and surveys once during the course of the study, dividing the study into 2 rounds.

Participants installed our VSCode extension on their system and cloned our study repository, then completed 6 annotation tasks focused on various code examples and annotation types, shown in Appendix \ref{sec:demo-examples}. Other than specific exceptions that we attribute to easily-addressed issues in current LLMs and LLM APIs (described in Appendix \ref{sec:user-study-lm-caveats}), all of the demo tasks were completed by round 2 participants without significant assistance from the interviewers.

\subsection{Need-finding Results}

Our need-finding pre-demo survey and discussion revealed several consistent pain points in managing code-related contextual information.

\subsubsection{Documentation Fragmentation}\label{sec:documentation-fragmentation}

Participants consistently described challenges with contextual information being scattered across multiple disconnected systems. P6 and P9 both discussed using shared and private drives for storing documentation, with P9 admitting that ``a really hard problem was that all of our docs were just like scattered in people's own Google drives.'' In other words, this useful documentation was effectively privately stored and inaccessible to other team members. P9 also summarized the consequences of this kind of fragmentation: ``[I]f you're looking at a piece of code and have a question about it, it's hard to know whether there exists a doc that would give some useful information.'' P6 echoed this, spontaneously saying that ``the documentation tends to be pretty far from the line by line parts of the code.'' Speaking about what they would want from documentation before seeing our system, P7 said that ``it would be nice at times to have some sort of link or embedded document, but writing out links in code is kind of ugly.'' This last point helps to explain why fragmentation happens: injecting external links or the documentation itself into code was perceived by participants as impacting the code's quality, as we expand on in Section \ref{sec:media-limitations}.

Users felt that \system{} did a good job of meeting the challenge of fragmentation. P6 felt that the organization of information through \system{} helpfully reduces the distance between relevant information, saying it would allow them to ``just go to the part of the code where the information I want is relevant, and then ... just go through this small set of comments that I've made over time.'' P8 called ``documentation ... that doesn't get lost in [their employer's] giant database of docs because it's attached to the code'' one of a number of ``killer features'' that \system{} could enable.

\subsubsection{Documentation Generation and Maintenance}

Participants generally recognized the importance of documentation, but pointed to practical challenges. Participants noted that documentation is often not prioritized or is ``undervalued'' in development workflows. Even P5, a professor who teaches software engineering, admitted skipping documentation when busy despite knowing ``the documentation will save me and others time and money in the future.'' Participants called initially creating documentation ``the hardest part'' of the documentation process and also consistently rated keeping documentation up to date with code changes as difficult.

While participants like P8 praised the system for enabling annotations to update and show continuous integration-like feedback as the code changes (see Section \ref{fig:lm-unit-test}), we noticed that the problem of initially creating documentation remained largely unaddressed. We discuss this limitation in Section \ref{sec:limitations-future-work}.

\subsubsection{Media Limitations}\label{sec:media-limitations}

Many participants expressed frustration with the limited expressiveness and forced trade-offs of current documentation tools. P6 complained that code comments ``constrain them to just words'' and said they use a physical notebook to draw things out, both for their own understanding and to communicate with other team members; they also later noted these notebooks eventually get thrown away (which makes sense based on Section \ref{sec:documentation-fragmentation}---the value of the notes drops as soon as it is hard to relate them to the code). Participants familiar with computational notebook systems wanted embedded images, rendered equations, and diagrams for code. P8 called this issue a ``pain point'' and described substitutes like ASCII art as ``a crutch, where you'd probably just want an SVG or a real image editing tool.''

As P8 put it, hosting ``arbitrary widgets that do cool stuff'' through our system could address this issue.

\subsection{Participant Responses to the Demo}

After interacting with our prototype, participants shared their reactions and insights about its utility and potential applications.

\subsubsection{Non-Intrusive Annotations Encourage Keeping More Context}

Most critically, \textit{all participants quickly understood the ability to maintain rich contextual information alongside code without modifying the source files as increasing possible context for a codebase}, usually bringing it up without the interviewer even mentioning the point. P6 described the system as letting them ``write down all these ideas and just have them around, but they're not imposing on the actual code.'' P3 saw the scalability advantage compared to inline comments: ``This is different from changing the code, because if everybody adds their comment to the code, then the code blows up with a huge amount of comments.'' P1 also described extended inline comments as ``annoying'' in comparison. P9 pointed to new opportunities to add extra information to codebases, like developers' provenance stories about how code got added to the codebase.



\subsubsection{Robust Code Tracking}

The system's ability to track annotations predictably as code is edited and through significant code changes was also valued by participants. P1 saw the advantage over traditional code comments, saying that ``when you're working with code, you don't want to have all those [traditional] comments everywhere because if you're moving stuff around, things get jumbled up.'' P3 imagined a system tracking the movement of annotations across files (which is not yet supported by our prototype) using a similar kind of semantic tracking. For the most part, it appeared that participants expected this aspect of the system to work out of the box, with P1 reacting to the LLM repositioning prompt after a complex update to the code by saying repositioning should happen ``in the background without me even having to think about it.''









\subsubsection{Interactive Features and Testing}\label{ssec:interactive-testing}

Several participants highlighted the potential of stateful interactive annotations. The Show Debugged Example annotation type shown in Figure \ref{fig:teaser} and described in Section \ref{sec:program-data-live-execution} was created based on recommendations from P2 and P3, who imagined annotations that show the data and execution process for code. P1 and P6 also recommended tools that permit a user to describe what they want a section of code to do at a high level and receive suggestions for how to do it; these recommendations inspired us to create the LM Unit Test annotation type we discuss in Section \ref{sec:lm-unit-test}. Our round 2 participants praised both of these annotation types, as we note in the relevant sections below.






\subsection{Future Vision}

In the final part of the post-demo discussion, participants shared their perspective on the role of systems like ours in the future of programming. P2 and P7 recognized the growing importance of connecting domain expertise with code, especially for applications in business and the sciences. P9 also envisioned \textit{annotations as facilitating a shift in programming abstraction and satisfying a new need for distinctions between human and model-generated information in documents}, speculating that
\begin{quote}
Maybe people will write pseudocode and then that will be linked in some way to the [model generated] code implementation of the pseudocode ... In that world, sorting through the noise will become harder... That seems like a place where annotation can be useful, like, here's a pile of code, and then here's some evidence that there's a person who has thought about it.
\end{quote}

Overall, our user study revealed significant challenges with current approaches to code documentation and strong potential interest in non-intrusive, robust, and interactive annotation systems like Codetations, especially as AI plays an increasing role in software development workflows.

\section{Applications}\label{sec:applications}

We now describe the annotation types we implemented or planned as test applications for \system{}. We targeted applications involving different roles for annotations and different interaction patterns between the user, the document, and the annotation. (More applications are described in Appendix \ref{sec:other-applications}.)
\begin{figure*}
    \centering
    \includegraphics[width=\linewidth]{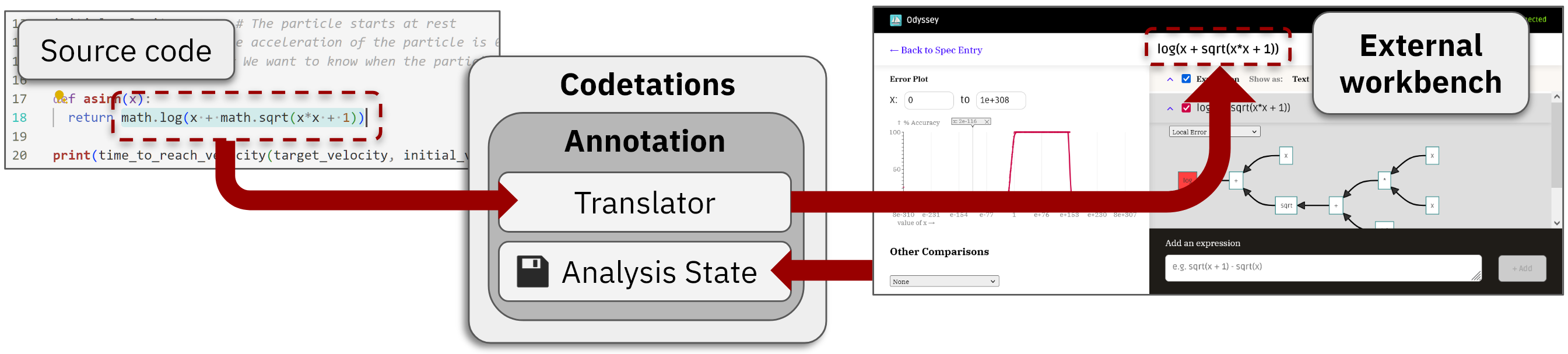}
    \caption{(right) \system{} connects the external Odyssey floating-point workbench, hosted on the web or in an editor, to (left) source code. An Analyze Floating-point Expression annotation uses an LLM to translate the Python expression shown on the left, math.log(x + math.sqrt(x*x + 1)), into log(x + sqrt(x*x + 1)), an expression in Odyssey's input language. This operation works for all input and output languages the LLM understands, immediately bringing Odyssey's analyses to a variety of languages that use floating-point operations. The user's work in Odyssey can be preserved in the annotation state.}
    \label{fig:codetations-odyssey}
\end{figure*}
\subsection{Program Data and Live Execution}\label{sec:program-data-live-execution}

Programming languages provide a high level of abstraction, helping developers handle varied cases with generic constructs. However, this abstraction complicates efforts to understand code behavior in specific instances or contexts. The Babylonian programming system of Rauch et al. \cite{rauch2019babylonian} solves this issue by annotating programs with example data to illustrate how code behaves with concrete inputs. Other systems that also implement ways of showing execution examples, such as interactive notebooks, resemble the system of Rauch et al. by generally requiring developers to adopt a particular development environment or toolchain when targeting production code.

Using \system{}, we rapidly developed the Show Debugged Example annotation type depicted in Figure \ref{fig:teaser} to assess whether it could address this problem (see Appendix~\ref{sec:show-debugged-example-prompt}). This annotation type lets users explore and record  the runtime behavior of any subsection of their program by attaching an annotation to it (for an example, see the summary and debugging code in Section 2B and execution results in Section 2C of Figure \ref{fig:teaser}). Users can write debugging code manually or use a language model to generate the code based on anchor text, file context, and user-provided high-level guidance (Section 2A of Figure \ref{fig:teaser}). If the file changes, users can ask an LLM to regenerate the debugging code while preserving as much of the existing debugging structure as possible (via the blue ``Regenerate'' button in Figure \ref{fig:teaser}).

Users in our study responded very positively to this annotation type. 
P2 observed, ``If you could 
run the code to get example output, that would...be useful for me,'' and P3 was enthusiastic about the possibility of both input and output value annotations for code (Section \ref{ssec:interactive-testing}). After seeing the tool we built, P9 noted that ``I didn't have to construct a whole input-output pair for some larger function. I could just say, `I want to test this line.' That's definitely useful.'' Thus, the increased \textit{addressability} of a \system{}-annotated codebase (the ability to work with individual code subsections, such as lines, as introduced by  Basman et al. \cite{basman2018open}) provides value relative to users' typical testing experience, partly by simplifying the testing process. P7 appreciated
that ``having [annotations with test cases] right next to the code is really nice.''

If given more access to a host's resources, the Show Debugged Example tool could run with different code execution engines to help programmers with any language. We could also foresee extending it with ``fuzzy execution,'' stepping through a program operation-by-operation using an LLM's notional machine rather than an engine, similar to a programmer carefully reading code during a debugging session to understand its function. Although a notional machine might initially be faulty and less reliable than a true engine, this is not necessarily a downside; we can view a debugging process as not just correcting the code, but also correcting the notional machine of the agent that generated it. Externalizing and aggregating critical corrections to a default notional machine over many annotations could ultimately be of lasting benefit to both programmers and models working with a particular codebase.

\subsection{Document Observers and Agents: Program Validation, Certification, and Summarization}\label{sec:agents}

In complex software systems, it is often desirable to consistently check code for certain properties. These properties might be formal proofs of behavior or dynamic checks for performance, or even developer-noted checks for whether the code meets some loose natural language specifications, like ensuring the documentation for a function is current. Validating these properties is vital to  ensure the correctness and performance of critical libraries and applications. Maintaining a historical record of these validations can facilitate monitoring the evolution of code quality and compliance over time. 

Unfortunately, these properties often describe code sections on the scale of individual expressions or loop bodies, where the developer would like to check for very specific execution concerns. This low-level logic is not readily exposed at the function level, so it is difficult to unit test.

\system{} enables the attachment of `document observer'  annotations directly to code snippets, solving the problem. Such observers could monitor high- or low-level properties and save their state over time. Below, we discuss several examples of observers.

\subsubsection{Analyzing floating-point expressions}\label{sec:floating-point-analyzer}
Floating-point expressions often suffer from inaccuracies that can propagate significant errors throughout a software application. Odyssey \cite{misback2023odyssey}, a web-based computational workbench, lets developers working with floating-point expressions explore and optimize expressions by rewriting them while viewing their error. Unfortunately,  such tools for analyzing floating point issues are scarce, and developers must locate and manually run them on particular expressions from a codebase even if their use is required for program validation. For library developers who rely on such checks to keep their libraries accurate and performing well, \system{} offers the  immediate benefit of bringing the tool to the code.

Integrating an external tool like Odyssey into diverse development environments requires a way to translate  between many possible input languages and the tool's input format. Odyssey expects expressions to be input in the MathJS format, which differs from the native formats used in most programming projects. Our ``Analyze Floating-point Expression'' annotation type finds a natural fit for this problem through the VSCode Language Model API, which we expose to annotations through the Annotation Type API in Codetations (see Section \ref{sec:annotation-types}). An extension could call this API to transpile arbitrary floating-point code into the MathJS format required by Odyssey, as shown in Figure \ref{fig:codetations-odyssey}.

This approach leverages Odyssey's human-readable input format, which is easy for a language model to produce. For popular programming languages, this translation streamlines the use of Odyssey and maintains a high degree of translation accuracy. Although more testing is needed, this approach demonstrates a clear pattern for integrating any tool with a well-defined input representation as a type of annotation. We envision the Analyze Floating-point Expression annotation type as just one example from an entire category of possible annotation types that wrap around existing analysis utilities with a well-defined intermediate representation used for input and output.

Study participants generally found this annotation type of interest for exposing a powerful analysis tool. According to P5, ``I loved the tool that helped me figure out how to improve my floating-point arithmetic. You don't just point out the problem or give a list of options. It gives me a framework for evaluating the options [accuracy vs performance trade-off analyses], which is great.'' Participants easily  discovered Odyssey and applied it to a codebase through our system, and future users of the same code could have found Odyssey output attached to the code to justify or certify their implementation choices. Significantly, if an implementation were to change, the same annotation would be immediately available to validate new choices.

\subsubsection{The LM Unit Test annotation type}\label{sec:lm-unit-test}

\begin{figure}
    \centering
    \includegraphics[width=\linewidth]{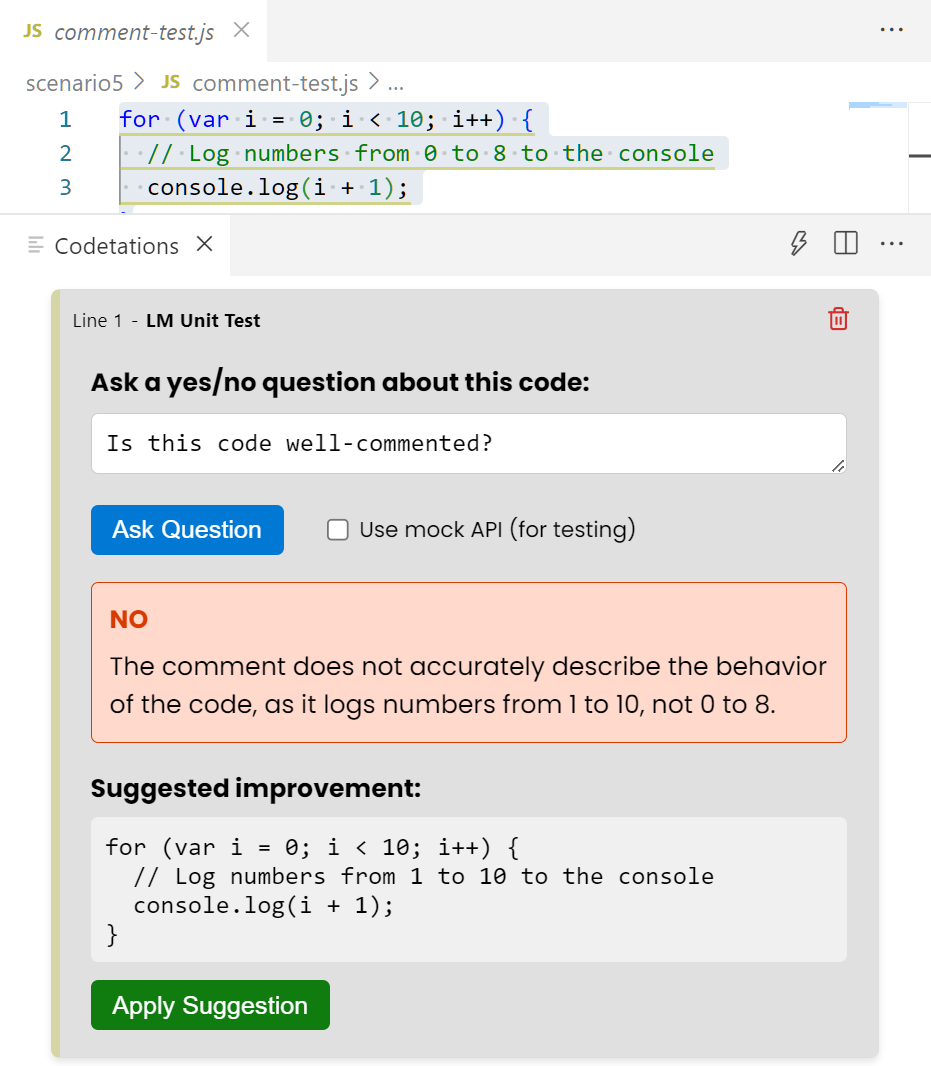}
    \caption{The LM Unit Test annotation type lets  users ask yes-or-no questions about the code and suggests changes if the answer is ``no.'' Suggestions can be applied to the buffer. The question stays with the code and can be asked again after the code is updated. This annotation type was generated in a few minutes by an LLM (see Appendix~\ref{sec:show-debugged-example-prompt}).}
    \label{fig:lm-unit-test}
\end{figure}

Consistently maintaining code documentation is a challenge in software development. Inline comments and documentation often become outdated or misaligned with  actual code logic due to updates and modifications in the codebase. This misalignment can lead to confusion, errors, and increased maintenance overhead.

By attaching an annotation that uses an LLM to observe code changes, we can dynamically check and update code sections for specific properties, including summaries or explanations. Our LM Unit Test, shown in Figure \ref{fig:lm-unit-test}, uses the editor's LLM API exposed by \system{} to provide a yes/no answer to a user's question about a code section; a `no' answer causes it to suggest updates that would change the answer to a yes. 

This annotation type has a variety of applications. Our study demonstrated the very simple case where a comment written by a beginner documenting the behavior of a loop is incorrect. The programmer might use the LM Unit Test in this case by asking if the loop is well-documented; the test will fail and suggest a correct comment. After the comment is corrected, the test will pass. P8 immediately perceived the value of this example: ``Now if someone changes this code, like, refactors it, but forgets to change the comments, which happens all the time, then this test will fail in CI.'' P9 mentioned a real-world analogue: ``Rust has mechanisms to run your documentation as tests... every piece of documentation should have a test embedded in it, and when that test fails, you go back and fix the documentation.''

Consider a more realistic scenario where a developer has written a method such that it has no side effects. To check that there are no unexpected side effects, they add an LM Unit Test that asks, ``Does this code have no side effects?'' to compare their understanding to the LLM's. Later, another programmer refactors the method to improve performance but accidentally adds side effects. The LM Unit Test annotation automatically detects these changes, warns the programmer, and suggests updates to the method to preserve the property of no side effects.

Our implementation of this annotation type requires the user to manually request the LLM check, but we believe that automation of this process would likely reduce the developer burden involved with manually checking documentation. Compared to asking an LLM if any comments in the codebase are out of date, this targeted documentation-checking approach would also likely increase the chance that specific known issues would be  noticed and thereby reach the developer's attention.

In general, annotation types like the LM Unit Test could reduce the manual effort required to keep documentation aligned with code changes, enhancing maintainability. We can envision similar annotation types that provide high-level summaries of the code they are attached to and manage the process of updating that documentation as the code evolves.



\subsubsection{Other application domains}

Keeping automatically-checked records with code has many other potential applications. On showing our system to a browser cryptography expert working on the Verified Software Toolchain project \cite{appel2011verified}, the problem of keeping proofs of subparts of cryptographic functions attached to the correct places in those functions as libraries evolve was immediately suggested, and the expert claimed that other critical systems like this also occasionally rely on manually updated attachments between certifications and certified code. While the risk of unpredictable output from an LLM's inclusion in a critical system should not be ignored, well-designed automated systems may make it much easier to identify certification issues as libraries evolve.

Live-updated visual representations of values that live next to the code, like the representation of floating-point expressions in Odyssey, are a basic annotation type that have already established their popularity in interactive notebook systems. When they are interactive, such representations can also become \textit{bidirectional} editors: changing the code changes the visual representation, and manipulating the visual representation leads to corresponding updates in the code. Hempel et al.'s Sketch-n-Sketch \cite{hempel2019sketch} exemplifies this technique. Bidirectional editors are a kind of \textit{structured editing} (first described as syntax-directed programming by \cite{teitelbaum1981cornell}), where a user is presented with an interface whose affordances let them generate valid programs or objects only by manipulating the program's semantic structure, not its text. Gobert and Beaudouin-Lafon \cite{gobert2023lorgnette} refer to these ``alternative interactive representations of specific fragments of code'' as \textit{projections}. Although purely syntax-based attachment tools (that match all instances of a syntax or regular expression pattern) may better fit some kinds of values, they cannot save state for individual projection instances (e.g., history or an undo chain) outside the code, and they might be augmented by a stateful \system{} data layer. \system{} also permits quick, intuitive use of the projection technique for larger or less well-defined code sections, as Odyssey demonstrates for floating-point expressions.

\subsection{Layered Documents}

In digital design tools like the popular Adobe Photoshop program, the concept of ``layers'' helps users manage multiple overlapping elements that contribute to a final composite image. However, this concept has not traditionally been applied to textual electronic documents or codebases, which present content as a single  consolidated file. The notion of layers in documents can offer a new way to separate and manage different types of content or functionality in a document.

Layered documents represent a novel user-document interaction pattern that a system like \system{} could  enable.


\subsubsection{The Add Layer annotation type}\label{sec:layered-documents}

\begin{figure*}
    \centering
    \includegraphics[width=1\linewidth]{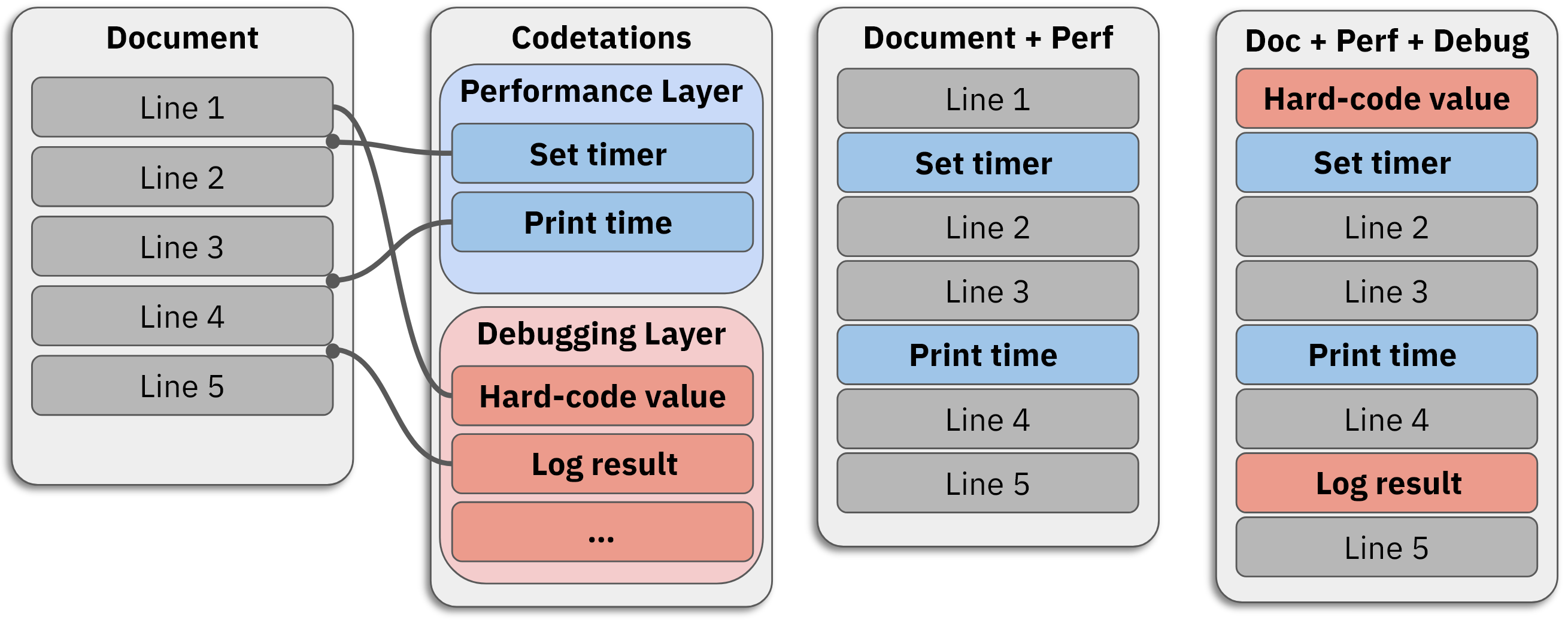}
    \caption{The document-external annotations enabled by our system could facilitate the separation of documents into layers that help with specific tasks, such as instrumenting code, without cluttering the main document.}
    \label{fig:layers}
\end{figure*}

Software development often involves auxiliary code segments that are necessary for tasks such as debugging, performance testing, or conditionally executed features (like feature flags). These segments, though useful during development or testing, can clutter the main codebase and complicate maintenance and readability. Using the concept of layers, we could overlay auxiliary code on a main codebase without permanently integrating it into the main files, keeping the primary code clean and focused.

Our proposed ``Add Layer'' annotation type (not implemented for our user study), shown in Figure \ref{fig:layers}, aims to enable developers to attach, manage, and toggle sets of auxiliary code segments as layers over their main codebase. The tool facilitates a clean separation of concerns by allowing developers to activate or deactivate different layers depending on the current need, such as toggling debugging annotations or performance metrics without altering the underlying source code.

To use the Add Layer annotation type, a developer would attach the annotation in the position where they would like to insert new code. Then, they would specify the code to be inserted at that position and the name of the layer to which the code belongs. Next, an ``ApplyLayers'' command-line tool could be used to read the codebase's annotation data files and obtain an alternate version of the codebase with a specified set of layers applied.

Consider a scenario where a developer needs to add extensive logging to troubleshoot a specific issue. Using the Add Layer type, they can create a new ``debugging'' layer where all logging code is added. This layer can be activated to gather necessary data and then easily deactivated once the issue is resolved, without leaving residual code in the production codebase. If the issue arises again after the code has changed, the layer can be reactivated to resume debugging. This approach keeps the codebase clean and formalizes the management of temporary or context-specific code modifications.

Layers could also be used by annotation types like the Show Debugged Example tool (Section \ref{sec:program-data-live-execution}) to automatically instrument code to gather runtime data.

\section{LLM Evaluation}\label{sec:machine-eval}

There's no question that context is required for any agent to resolve issues in a domain it knows nothing at all about. However, considering the amount of training data shown to modern LLMs and advances in their intelligence, is context still necessary for a domain like program repair? To investigate RQ3 and explore the impact of context on LLM reasoning (see Section \ref{sec:research-questions}), besides consulting the work of authors like Martino et al. on hallucination \cite{martino2023knowledge}, we also evaluated a few examples to understand to what extent context is necessary for current LLMs (ca. 03/2025) to recall and apply knowledge \textit{known to be latent in the model}, especially for problems involving program repair. We also wanted to explore the effect of irrelevant information.

To guarantee that the information was latent in the model, we began by generating example debugging problems with an LLM without tool calling. Each debugging problem consisted of a program requiring repair and an explanation of the bug. Next, in each case, we constructed context that simulates real-world cues: partial documentation, debug outputs, and other JSON-formatted annotations based on those output by Codetations, with a hint at the true explanation mixed into a set of extraneous information. These sets of contextual information mimic the kinds of fragmentary artifacts one might find on a voluntarily-annotated codebase. Finally, we observed output for \textit{the same} LLM when prompted to solve each debugging problem (1) without and (2) with context.

Our null hypothesis was that even in the zero-context condition, an LLM with excellent recall might be able to guess solutions, since it was able to think of the problems. If this were the case, the value of keeping particular notes on the code for problems an LLM knows about would diminish. However, our results support the view that, for current LLMs at least, even when extraneous information is included, useful context still significantly improves performance, particularly in scenarios where failures stem from missing external references or real-world assumptions. The following section examines one failure mode in detail, with a second case shown in Appendix \ref{sec:worked-examples}.

\subsection{External References}\label{ssec:extref}

LLMs, like human programmers, rely on external references, such as inline comments, documentation, or APIs, to bridge gaps in their internal knowledge. When these references are omitted or misremembered, both humans and LLMs make predictable errors. 

Consider a bug involving the GitHub API: the endpoint https://api.github.com/users/\{username\}/repos returns only public repositories unless an authentication token is supplied but returns 200 OK in both authenticated and unauthenticated cases. We prompted GPT-4o with this setup and a description of the bug: ``This code returns no repositories for some users.'' Without context, the model recognized that the issue was related to access control but misattributed it to using an incorrect endpoint (and switched the code to using a different, incorrect endpoint). In reality, the endpoint was correct: the problem was missing authentication.

When we provided annotations, including distractor documentation and data from a Show Debugged Example annotation demonstrating the unauthenticated response, the model immediately diagnosed the issue and inserted the correct token logic. However, we also note that when the model was prompted with only unrelated context, such as documentation about pagination limits, it confidently produced incorrect diagnoses \textit{citing the unrelated context}.

This simple case shows that a system intending to leverage context to help LLMs should consider the risk of unrelated context being abused by the model to create convincing model hallucinations, especially in the case that a correct solution is missing from the context. Determining relevance is likely much more challenging than simple ``needle-in-a-haystack'' retrieval, since many statements might falsely seem relevant. Proper benchmarks around this challenge are called for, with the assignment of probabilities for the relevance of individual items of context as a target.

Overall, this reflects a broader phenomenon: LLMs, like human developers, benefit from context that simulates embedded team knowledge, and shared representations are valuable; however, models can also be led astray in ways that humans usually are not, and humans also have their own limitations. While sharing context may benefit both humans and LLMs, systems like \system{} may need additional actor-specialized scaffolds around the use of that context to safely support different kinds of actors in a codebase.

\section{Discussion}\label{sec:discussion}

This section reflects on what we learned from building and studying \system{} and offers design takeaways for researchers and tool builders who want to treat annotations as rich and evolving artifacts rather than static comments.


\subsection{Reflections on our Research Questions}
\label{sec:rq-discussion}

\subsubsection{RQ1: Utility of semantically anchored, interactive notes.}
Hands-on tasks showed that developers valued (i) robust re-anchoring during edits, (ii) live executable widgets such as the Show Debugged Example annotation type, and (iii) flexible testing tools like LM Unit Test.  
Participants called documentation that stays next to related code a ``killer feature'' and could think of use cases in their own work for the interactive components we showed.

\subsubsection{RQ2: Evolving contextual needs.}
Interviews revealed that developers struggle with \textit{documentation fragmentation}, \textit{maintenance effort}, and the \textit{expressiveness limits} of plain comments.  
Participants pointed to missing links between code and scattered artefacts while expressing concern about documentation methods that increase code clutter.
For future programming systems, they envisioned an increased need for links between code and high-level documentation.

\subsubsection{RQ3: Impact of context on LLM reasoning.}
Our controlled experiments demonstrate that GPT-4o solved repair tasks reliably only when the relevant slice of annotation context was present; without it, the model offered incorrect fixes despite having the knowledge in its parameters.  
Thus, curated, local context still materially boosts state-of-the-art models and can guard against confident error, but irrelevant context can also mislead models. This highlights the need for both context filtering and provenance tracking.

\subsection{Other Key Design Takeaways}
\textbf{Direct integration beats side‑car tooling.} Pilot users struggled with an early browser prototype located outside their editor; embedding the system as a VSCode extension immediately improved comprehension and adoption. The lesson echoes decades of IDE research: if an augmentation must be consulted continuously, it needs to share the developer’s focal surface.

\textbf{Rapid creation is practical.} Multiple powerful annotation types were generated with a single prompt once the API was demonstrated, confirming that modern LLMs make the long‑standing vision of end‑user‐programmable IDE widgets realistic. Practically, this means that a small team can seed a rich ecosystem; the barrier is no longer engineering effort,  but surfacing needs.

\subsection{Limitations and Future Work}\label{sec:limitations-future-work}

Our evaluation relied on a small convenience sample of nine participants drawn mainly from research environments. Industry teams with entrenched workflows may surface different pain points or adoption barriers (or offer faster paths to adoption). 

The system also assumes that developers will invest effort to create annotations, although \textbf{users told us that time is their chief obstacle}. Our system might make adding context a more natural part of the development process, but using it is still far from passive; users must select text and interact with the UI to provide context. We think that addressing users' documentation fatigue will probably require a more automated approach that collects information passively as part of a normal development workflow, or even actively prompts the user (without being intrusive) to collect context on decisions.

Participants also asked for many additional features that we did not have time to build, such as annotations that span multiple code regions, possibly across files, or annotations that apply to the whole file. These would require changes in our annotation definition, but they seem otherwise feasible in the system we described.

Finally, our implementation of \system{} remains a prototype, and revision and more extensive user testing, especially with teams of users, is required to fully understand the impact of a \system{}-like  system on real-world development workflows.

\section{Conclusion}
\system{} demonstrates that lightweight, LLM‑aware annotations can externalize rich context, survive code evolution, and even act on their surroundings without modifying source files. A VSCode prototype, a qualitative user study, and targeted LLM experiments show that the approach is both feasible and valued by developers. The path forward lies in reducing the effort to add notes, scaling anchoring across languages, and exploring how active annotations can mediate ever closer human-LLM collaboration.


\bibliographystyle{ACM-Reference-Format}
\bibliography{codetations}

\appendix

\section{Basic Annotation Affordances}\label{sec:basic-annotations}

\subsection{Connecting the buffer and annotation view}
When the cursor enters the region of some anchor text, an associated annotation is shown and highlighted in the annotation view; if an annotation is clicked on, the editor scrolls to the related anchor text and highlights it. Annotations are also assigned customizable colors that outline their left edge and underline the anchor text. They are also marked with their line number.

\subsection{Adding and moving annotations}
New annotations are added by selecting text in the file and then, in a dialog that appears in the annotation view during text selection, selecting an annotation type and clicking a button to confirm. A selected annotation can be moved to another section of the text by selecting the annotation, selecting the new anchor text, and clicking on the ``Move'' button in the selected annotation or on the editor's context menu.

\section{Extra Implementation Details}\label{sec:implementation-extra}



\subsection{Observing edit actions}
As a VSCode extension, our host process responds immediately to user edits of the buffer by updating relevant annotation positions, as described in Section \ref{sec:location-updates}. An earlier implementation used a WebSocket-based document server to monitor the disk for changes, but we found that the latency on annotation position updates using this method---essentially, waiting for the user or editor to save the document before updating underlined anchor text---was confusing for users, who expected annotations to move as they typed.

\section{Show Debugged Example Annotation Type Generation Prompts}\label{sec:show-debugged-example-prompt}

Here we show prompts used to rapidly generate the tool shown in Figure \ref{fig:teaser}.

The prompts were given to Claude 3.7 via GitHub Copilot Chat in VSCode, along with the code for instantiating 5 previously-existing annotation types.

1: ``Add a Show Debugged Example annotation type that attaches to javascript code and asks a language model to produce code that runs an example execution of the selected region (with any missing variables set to reasonable defaults) with debugging instrumentation. The produced code should be held in a cached user-editable field and should execute (for now just using the webview interpreter) and print results visibly for inspection when the user clicks a button. In order to make sure the produced code uses reasonable defaults you should use props.value.document to inform the language model about code context (the version of the code cached in the annotation), as in LMUnitTest.''

This produced a working component with all of the fields shown in Figure \ref{fig:teaser} except for the natural language end-user customization field at the top of the component.

2: ``Update this component to include a field that allows a user to adjust what is debugged with a natural language prompt.''

This produced a working component with all of the fields shown in Figure \ref{fig:teaser}.

3: ``When regenerating, the debugging code may already be populated. Any existing debugging instrumentation and any examples used should be preserved if possible.''

This produced a working component with the underlying generation prompt updated to include previous context.

\begin{figure}
    \centering
    \fbox{\includegraphics[width=0.95\linewidth]{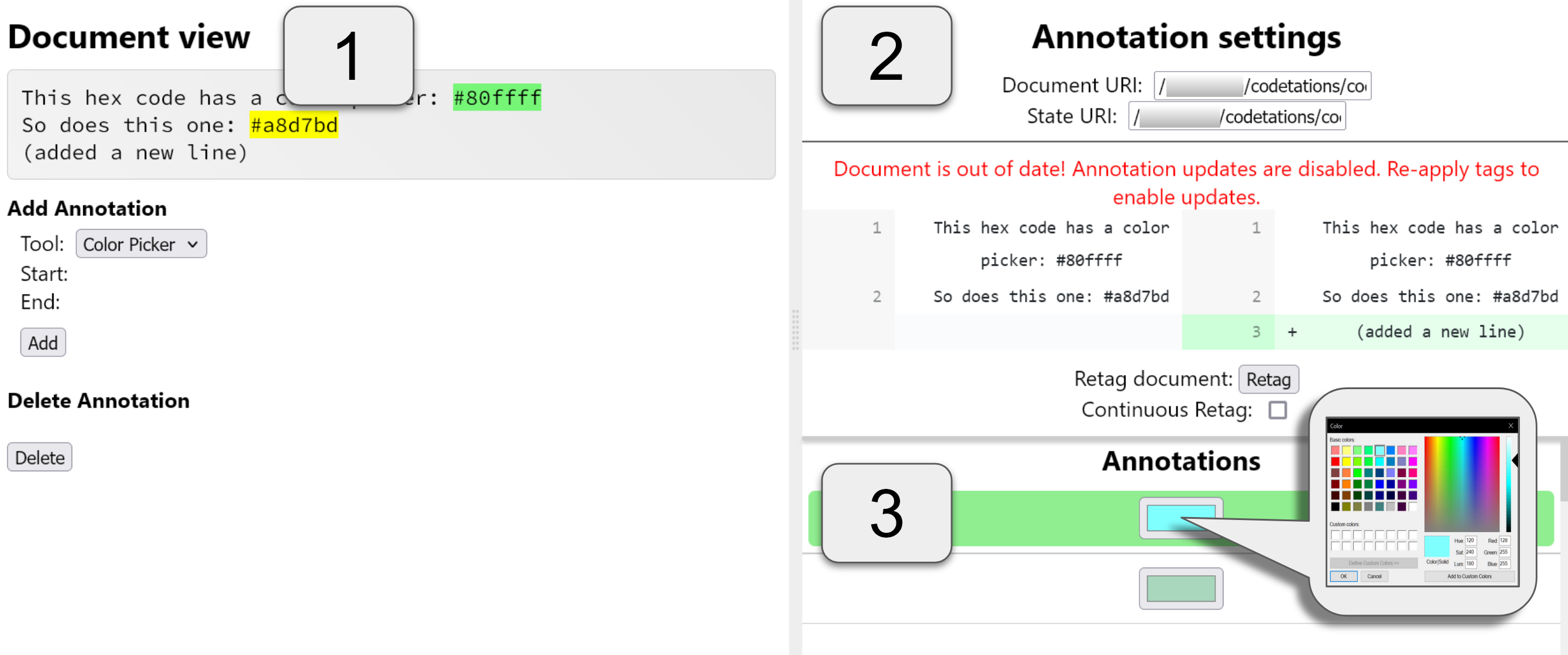}}
    \caption{An early prototype of our system was fully editor-independent and ran in a browser. It featured (1) a document view to allow users to select annotation targets, (2) an annotation view with (user-irrelevant) configuration and document attachment information, and (3) a list of annotations. Even after explaining the system's function, users struggled to imagine workflows involving file annotations hosted in a browser separate from their editor.}
    \label{fig:pilot-ui}
\end{figure}

\section{LLM-Generated Code Challenges}

\subsection{GitHub API}

\textbf{Summary:} As discussed in Section~\ref{ssec:extref}, this example demonstrates how external references can significantly improve a language model’s reasoning about code. The GitHub API endpoint for listing a user's repositories returns only public repositories unless an authentication token is provided. When presented with Python code that performs an unauthenticated request, language models frequently suggest using a different endpoint—misidentifying the source of the problem. However, when supplemented with relevant API documentation (alongside some extraneous, distractor context), the models accurately and efficiently diagnose the issue, correctly identifying missing authentication as the root cause.

\textbf{Example code:}
\begin{verbatim}
import requests

def list_repos(username):
    url = f"https://api.github.com/users/{username}/
        repos"
    response = requests.get(url)
    if response.status_code == 200:
        return response.json()
    else:
        raise Exception("Failed to list repositories")
\end{verbatim}

\textbf{LLM Prompt:} Help debug this code. It returns no repositories for some users.

\textbf{Context:}

\begin{itemize}
    \item GitHub’s REST API endpoint \texttt{/users/:username/repos} only returns public repositories. Access to private repositories requires authentication via a personal access token with \texttt{repo} scope.
    \item API responses are paginated. By default, only 30 repositories are returned. Use \texttt{?per\_page=} and pagination headers to retrieve more.
    \item GitHub usernames are case-insensitive, may include alphanumeric characters or hyphens, and cannot begin or end with a hyphen.
    \item GraphQL is also available in GitHub’s API for more efficient and flexible queries.
    \item Code before and after debugging:
\begin{verbatim}
# Original version:
def list_repos(username):
    url = f"https://api.github.com/users/
        {username}/repos"
    response = requests.get(url)
    if response.status_code == 200:
        return response.json()
    else:
        raise Exception("Failed to list 
        repositories")

# Debugged version:
def list_repos(username, token=None):
    url = f"https://api.github.com/users/
        {username}/repos"
    headers = {}
    if token:
        headers['Authorization'] = f'token 
            {token}'
    response = requests.get(url, headers=headers)
    data = response.json()
    if response.status_code == 200:
        if isinstance(data, dict) and 'message' in 
            data:
            raise Exception(f"GitHub API error: 
            {data['message']}")
        return data
    else:
        raise Exception("Failed to list 
        repositories")
\end{verbatim}
    \item Example input: \texttt{username = "octocat"}
    \item Output (before): public repos only
    \item Output (after): public and private repos (with token)
\end{itemize}

\textbf{LLM Response Without Context:}

Language models usually suggest using a different GitHub API endpoint or point out that the 200 result could signal a ``partial success.''

\textbf{LLM Response With Context:}

Language models typically point out how authentication is required and add boilerplate code to provide the authentication.

\subsection{NFC Chip with Adafruit}

\textbf{Summary:} As mentioned in Section ~\ref{ssec:realworld}, hardware-specific constraints often reduce LLM performance when generating or reasoning about code. The PN532 NFC reader, when configured in UART mode, requires an ACK byte after every frame. When given Python code that initializes the reader in this mode without handling ACKs, language models frequently misdiagnose the issue—often inserting unnecessary sleep delays under the assumption that the problem lies in timing. In reality, the correct fix is to either explicitly handle the ACKs or switch to high-speed UART mode, which removes the requirement altogether. When provided with PN532 documentation, models reliably identify the correct resolution. Notably, some reasoning-augmented models are able to infer the need for ACK handling even without additional context.

\textbf{Example code:}
\begin{verbatim}
def read_nfc_tags(nfc_uart):
    nfc_uart.write(b'\x00\x00\xFF\x02
        \xFE\xD4\x4A\x01\x00
        \xE1\x00')  # InListPassiveTarget
    response = nfc_uart.read(16)
    return response
\end{verbatim}

\textbf{LLM Prompt:} Users report that sometimes the NFC reader returns partial or no data when attempting to scan tags. This occurs inconsistently, and restarting the module temporarily fixes it. Please debug this function.

\textbf{Context:}
\begin{itemize}
    \item The PN532 supports multiple host interfaces including I2C, SPI, and UART. It is compliant with ISO/IEC 14443 Type A/B and FeliCa and supports peer-to-peer communication using NFC Forum standards.
    \item According to the PN532 user manual (UM0701), in Normal UART mode the host must send an ACK byte after every frame. If ACKs are missing, the chip enters a discard-and-recover state. Switching to HSU mode disables this requirement and enables continuous polling without dropped frames.
    \item Code comparison:
\begin{verbatim}
# Without mode switch (normal UART mode):
nfc = PN532UART()
nfc.write(b'\x00\x00\xFF\x02\xFE\xD4
    \x4A\x01\x00\xE1\x00')
print(nfc.read(16))

# With HSU mode enabled:
nfc.reset()
nfc.write(b'\x00\x00\xFF\x02\xFE\xD4
    \x14\x01\x17\x00')  # HSU switch command
nfc.set_baud(115200)
nfc.write(b'\x00\x00\xFF\x02\xFE
    \xD4\x4A\x01\x00\xE1\x00')
print(nfc.read(16))
\end{verbatim}
    \item Response comparison:
        \begin{itemize}
            \item Without mode switch: \texttt{[]}
            \item With mode switch: \texttt{[0x00, 0x00, 0xFF, 0x03, 0xFD, 0xD5, 0x4B, 0x01, 0x00, ...]}
        \end{itemize}
\end{itemize}

\textbf{LLM Response Without Context:}

Usually, OpenAI 4o and Claude 3.5 Sonnet suggest adding a short waiting period after every 16 bytes. Reasoning models suggest sending an ACK byte after each read.

\textbf{LLM Response With Context:}

Most types of language models switch the module to high-speed UART mode on initialization, which would solve the problem as an ACK byte is not required anymore.

\section{Annotation Types Participants Wanted}

\subsection*{Testing and Execution}
\begin{itemize}
\item \textbf{Test Generation}: P9 was ``most likely to want model generated code for'' test generation, asking ``Is my test coverage sufficient? What edge cases am I missing?''
\item \textbf{Inline Tests}: P2, P7, and P3 all wanted inline test capabilities. P7 specifically requested ``tests that are just in line... you can like run the test.''
\item \textbf{Input/Output Examples}: P3 wanted to ``show in quotation, like, what's this input? This is the output.''
\item \textbf{Execution Results}: P2 requested ``If you could do stuff where you could run the code to get example output, that would also be useful for me.''
\item \textbf{Conditional Execution Tracking}: P9 suggested ``Some way to have an annotation that's conditionally executing parts of your code and tracking the results for many different configurations.''
\end{itemize}

\subsection*{Visual Content}
\begin{itemize}
\item \textbf{Images}: P1 wanted ``Comments, images, have a snapshot of what the code running would look like.''
\item \textbf{Diagrams}: P6 mentioned ``I could put in diagrams and stuff.''
\item \textbf{Mathematical Formulas}: P7 specifically requested mathematical formula visualization: ``some annotation that you highlight a line of code and indicate this is the mathematical formula that this is implementing''
\item \textbf{LaTeX formulas}: P3 wanted to ensure ``latex format formula matches what you prove in the code''
\item \textbf{Data Structure Visualizations}: P8 desired visualizations for data structures: ``if every single cargo test I had that involved an egraph had a quotation that let me explore the egraph and interact with it''
\item \textbf{Tables}: P4 requested both ``Rich text editor - insert tables'' and ``Rendering of the table that's produced by the code''
\end{itemize}

\subsection*{Documentation and References}
\begin{itemize}
\item \textbf{Document-wide Annotations}: P2 suggested ``Instead of having it on just some texts, you could just add like a document annotation''
\item \textbf{Ownership Information}: P2 wanted ``information that's document wide'' including ``who owns the code''
\item \textbf{Research Paper Links}: P7 requested embedding research papers: ``annotate the code in such a way that I can embed some... diagram or a paper or the entire paper, a snippet of the paper''
\item \textbf{Data Provenance}: P7 wanted ``annotations that would be sort of provenance, both in terms of what experiment loads this data, and also... where did I get these constants?''
\end{itemize}

\subsection*{AI-Powered Features}
\begin{itemize}
\item \textbf{AI Suggestions}: P1 thought ``AI suggestions would be cool—if there's something confusing, you hover over it, and it gives you a suggestion.''
\item \textbf{Code Refactoring}: P2 suggested managing code refactoring through an annotation.
\item \textbf{Intent-Based Code Fixing}: P6 suggested AI could ``see what [the code is] doing now and read about what you want it to do and just tweak it''
\item \textbf{Pseudocode Links}: P9 envisioned ``pseudocode and then that will be linked in some way to the code implementation''
\end{itemize}

\subsection*{Code Quality and Analysis}
\begin{itemize}
\item \textbf{Performance/Optimization Comments}: P3 mentioned needing to explain ``optimized code vs. easy-to-read code''
\item \textbf{Coherency Checks}: P3 wanted ``Coherency between machine proof and pen-and-paper''
\item \textbf{Live Preview}: P4 praised and wanted more HTML preview functionality with live rendering capabilities
\item \textbf{Side-by-side Comparisons}: P8 wanted ``the original file and the produced output side by side in VS Code''
\end{itemize}

\begin{table*}
  \caption{Participant backgrounds for our user study.}
  \label{tab:participant-backgrounds}
  \begin{tabular}{llll}
    \toprule
    ID&Round&Background&Prog. Exp.\\
    \midrule
    P1 & 1 & Undergrad, maintains a documentation generation extension & 5 years\\
    P2 & 1 & Senior Research Specialist, chem. engr. and synth. bio. & 10 years\\
    P3 & 1 & Postdoc, proof engineering and system programming & 18 years \\
    P4 & 1 & MBA, SQL development & 25 years\\
    P5 & 2 & Senior Lecturer, user-facing formal methods and ed.; tax software & 30 years\\
    P6 & 1 & Undergrad RA, DSL/IR research  & 7 years\\
    P7 & 2 & Grad. RA, soft. engr. \& PL for computing ed.; comp. synth. bio and robotics & 12 years\\
    P8 & 2 & Grad. RA + intern, compilers, verification; game dev. & 9 years\\
    P9 & 2 & Grad. RA + intern, compilers, verif.; maps app, cloud security, block prog. & 10 years\\
  \bottomrule
\end{tabular}
\end{table*}

\begin{table}
  \caption{Quantitative responses from round 2 of our user study. (*P7 works with computational notebooks and reported less trouble with maintaining documentation.) Our round 2 pre- and post-surveys are shown in Appendices \ref{sec:pre-survey} and \ref{sec:post-survey}.}
  \label{tab:round-2-quantitative}
  \begin{tabular}{llll}
    \toprule
    ID & Difficulty of & Difficulty of & Likelihood of \\ 
    & Doc. (1-7)&Doc. Sync. (1-7)&Using Our Tool (1-7)\\
    \midrule
    P5 & 6 & 7 & 5 \\
    P7* & 2 & 4 & 6 \\
    P8 & 6 & 5 & 7 \\
    P9 & 6 & 7 & 6 \\
  \bottomrule
\end{tabular}
\end{table}

\section{LLM-related User Study Issues}\label{sec:user-study-lm-caveats}
One participant (P5) was unable to activate GitHub Copilot, which is currently required for the VSCode Language Model API that backs several features of our system, including the offline annotation position updates. This highlights a need to include basic free or local alternative APIs as fallbacks in a full system. We note that P5 was still able to complete parts 1, 2, and 4 of the demonstration, showing that the basic system and some annotations can still operate in the absence of a language model.

We also note that the offline position update using a language model\footnote{GPT-4 Turbo and GPT-4o during our study} failed a couple of times during our study for the regular expression annotation. While this was unfortunate, it wasn't unexpected: Misback et al.~\cite{Misback2024MagicMM} note that current language models still struggle with outputting exact copies of some kinds of text, and they acknowledge that their position update system could stand further improvement in that it requires the model to output text that matches the new annotated section exactly. We did not spend study time innovating on that basic method, since it has already improved during the course of our study with improvements in the underlying models and could be further boosted with a properly-implemented fuzzy match. We did observe that the participants presented with this failure asked if they could move the annotation manually, and were able to do so, indicating again that the system's ability to guarantee basic annotation functions even in the absence of a language model has value.

Overall, we were satisfied that basic annotation features could continue operating even when bleeding-edge features fail. While we think it's important to address these issues in future versions of Codetations, we don't view them as insurmountable problems for the idea of semantic annotation attachment.

\section{User Study: Significant Quotes}

\subsection{User-Identified Needs}

Our interviews revealed several consistent pain points in managing code-related contextual information:

\subsubsection{Documentation Fragmentation}

Participants consistently described challenges with contextual information being scattered across multiple disconnected systems:

\begin{quote}
``Documentation in a shared drive, the documentation tends to be pretty far from the line by line parts of the code.'' (P6)
\end{quote}

\begin{quote}
``I don't think there's any way for a person looking at my code to know, `oh, there's this Excel or Word document that you should be looking at.''' (P4)
\end{quote}

This disconnection creates significant barriers to code comprehension. P9 elaborated:

\begin{quote}
``All of our docs were just like scattered in people's own Google drives. And so if you're looking at a piece of code and had a question about it, it was hard to know whether there existed a doc that would give some useful information.'' (P9)
\end{quote}

\subsubsection{Documentation Generation and Maintenance}

Participants overwhelmingly rated keeping documentation synchronized with code changes as fairly difficult, with most rating it 5-7 on a 7-point scale:

\begin{quote}
``Look, when you have a deadline, what's more important: the code or the documentation? I teach software engineering, and I struggle with this... Surely I'll do the documentation afterward.... Spoiler: I won't; there's another crisis.'' (P5)
\end{quote}

P2 summarized: ``The hardest part is just doing it. And just being consistent with it, and taking the time to do it.'' Several participants noted that documentation is often not prioritized in development workflows:

\begin{quote}
``I think there are too many demands on developer time and documentation tends to be undervalued by developer teams... it's theoretically my job, but like not actually given time in the sprint plan... We'll ship the feature without it. And so we never come around and get back to the documentation.'' (P9)
\end{quote}

\subsubsection{Media Limitations}

Many participants expressed frustration with the limited expressiveness of current documentation tools:

\begin{quote}
``[Asked about problems with code comments as documentation:] I like to use the notebook [a physical notebook], because I can draw things out other than words, which I think is very helpful to understand what's going on and to communicate what's going on. So being constrained to just words.'' (P6)
\end{quote}

P2 desired embedded visualizations: ``If there were a way to just embed [an image] into the code... like, here's it before the function, here's it after the function, that would be useful.'' P9 observed that developers attempt workarounds with ASCII diagrams, but the effort required discourages wider use:

\begin{quote}
``I've seen people do like kind of ASCII art level diagrams as comments... I think that kind of thing would be more widely done if it were easier. Like if you could not do it as ASCII art... because ASCII art is kind of a pain... people only do it if it's like really complicated.'' (P9)
\end{quote}

\subsection{User Responses to the System}

After interacting with our prototype, participants shared their reactions and insights about its utility and potential applications.

\subsubsection{Value of Non-Intrusive Annotations}

Participants appreciated maintaining rich contextual information alongside code without modifying the source files:

\begin{quote}
``I feel like I could, in [Codetations], kind of like write down all these ideas and just have them around, but they're not imposing on the actual code.'' (P6)
\end{quote}

P3 recognized the scalability advantage of external annotations compared to inline comments:

\begin{quote}
``This is different from changing the code, because if everybody adds their comment to the code, then the code blows up with a huge amount of comments, and that's not going to be very helpful.'' (P3)
\end{quote}

P9 reflected on the distinction between inline comments and annotations:

\begin{quote}
``A code comment feels more like it should be about the syntax as it stands in some way, like the code, not about the functionality... And then we'd want to use an annotation [in \system{}] to say something like, `we did this for this product launch on this day,' things that are a little higher level than the code itself.'' (P9)
\end{quote}

\subsubsection{Robust Code Tracking}

The system's ability to track annotations through significant code changes using reattachment based on document semantics was valued:

\begin{quote}
``If a function is moved from one file to the other, then I would fail in trying to go through the old documentation I have. But I think for this one, if there is retag, the higher chance I will be able to do that.'' (P3)\footnote{Note the participant here was imagining a future feature of our system, which only handles re-tagging within a file at present.}
\end{quote}

P1 noted the potential benefit of automatic retagging: ``That would be nice if it just does that in the background without me even having to think about it.''

\subsubsection{Utility for Different User Roles}

Participants identified value for various roles within software development teams:

\begin{quote}
``For the original code author, it does help [them] to be more aware of... to be able to communicate more efficiently about the intent of the code.'' (P3)
\end{quote}

\begin{quote}
``For the code maintainer... it's easier for them to get that code context information.'' (P3)
\end{quote}

And for code testing:
\begin{quote}
``It would be nice to see which part of the code works or doesn't work... [it] would help with the QA process.'' (P4)
\end{quote}

\subsubsection{Interactive Features and Testing}

Several other participants highlighted the potential of the system's interactive capabilities.

P3's idea for illustrating code behavior were later adapted into the Show Debugged Example tool shown in Figure \ref{fig:teaser}:

\begin{quote}
``It would be immensely intuitive if I can just show in \system{}, like, what's this input? This is the output.'' (P3)
\end{quote}

P9 later showed excitement at the prospect of debugging with the Show Debugged Example tool:

\begin{quote}
``I think the test generation is like the thing that for me, I would be most likely to want model generated code for... generate some test inputs for this. Is my test coverage sufficient? What edge cases am I missing?'' (P9)
\end{quote}

P8 contemplated using annotations for in-context exploration of data structures: ``I do work with a lot of data structures that have visualizations. So if every single cargo test I had that involved an e-graph had [an annotation] that let me explore the e-graph and interact with it or build new tests, that would be pretty cool.''

\subsection{Future Vision}

Participants shared their perspectives on how systems like Codetations might shape the future of programming.

\subsubsection{AI and Annotation Integration}

Many participants envisioned deeper integration between annotations and AI assistance:

\begin{quote}
``Maybe if you knew that a part of your code wasn't working how you wanted it to you could comment on it and say like this is what I describe what you want it to be doing and then the AI could like put the code into the context of the system... see what it's doing now and read about what you want it to do and just tweak it.'' (P6)
\end{quote}

P9 also envisioned a shift in programming abstraction:

\begin{quote}
``Maybe people will write pseudocode and then that will be linked in some way to the code implementation of the pseudocode... the model is going to generate the actual code that runs in production. But the thing that we're all looking at and talking about is a pseudocode model of it. And then we need links between the various points.'' (P9)
\end{quote}

\subsubsection{Domain Knowledge Representation}

Participants recognized the growing importance of connecting domain expertise with code:

\begin{quote}
``I think that as programming becomes more connected to `real world' tasks outside of pieces of code---for example by scientists, social scientists, businesses---it becomes more and more important to include the contextual information and domain knowledge related to programs with the programs themselves.'' (P7)
\end{quote}

P9 noted the specific value for AI-generated code:

\begin{quote}
``I think the future of programming includes more and more computer generated code... in that world, sorting through the noise will become harder if you're just like more and more low to medium quality code, or like, at least unknowable quality code... that seems like a place where annotation can be useful is like, here's a pile of code. And then here's some evidence that there's a person who has thought about it.'' (P9)
\end{quote}

\section{Pre-demo Surveys}\label{sec:pre-survey}
Participants were asked the following questions in our pre-demo survey. The questions mainly served as a starting point for discussion, and most participants skipped one or more questions due to time constraints.

\subsection{Round 1}
Question: What is your education level and job title (if applicable)?

Question: How many years of programming experience do you have?

Question: Briefly, in what kinds of positions or research domains have you worked as a programmer?

Question: What is your primary development environment (e.g., VSCode, IntelliJ, other)?

Question: Do you have any experience using Copilot or another code generation/programming assistance platform?

Question: How do you currently maintain code documentation?

Question: When working with a team, how do you handle sharing and storing contextual information about code?

Question: How do you currently handle attaching supplementary information (like examples, visualizations, or analysis results) to specific sections of code?

Question: What kinds of problems have you faced with code documentation systems? For example, have you faced challenges in keeping documentation synchronized with code changes?

\subsection{Round 2}
Question: Education level + job titles if applicable

Question: Years of programming experience

Question: Briefly, in what kinds of positions or research domains have you worked as a programmer?

Question: Primary development environment (VSCode/IntelliJ/other):

Question: Do you have any experience using Copilot or another code generation/programming assistance platform? If so, when prompting a language model to assist with your work, what kinds of information do you usually provide?

Question: Please skim this list of types and examples of contextual information that developers may use while working with code [Appendix \ref{sec:types-of-dev-context}]. Briefly, how do you currently maintain code documentation and contextual information for projects that you work on?

Question: On a scale of 1-7, in your experience, how difficult is it in practice to write documentation while programming?

Question: On a scale of 1-7, in your experience, how difficult is it to keep documentation synchronized with code changes over time?

Question: What specific problems have you faced with maintaining code documentation and contextual information?

Question: Describe a situation where better code annotations would have been valuable to you or your team.

Question: Do you currently attach long-form contextual information (e.g., examples, tutorials, notes, diagrams, or videos) to specific code sections? If so, how? If not, why not?

\section{Types of Development Context}\label{sec:types-of-dev-context}
The following lists were provided to participants in round 2 of our user study to help them reflect on their current context maintenance processes.

\begin{itemize}
  \item \textbf{General types}
  \begin{itemize}
    \item \textbf{Design \& Architecture Information}
    \begin{itemize}
      \item Architecture diagrams and system flow charts
      \item Design patterns used and their rationale
      \item Component relationships and dependencies
      \item System constraints and design trade-offs
      \item API contracts and interface specifications
      \item Architectural decisions and their justifications
    \end{itemize}
    
    \item \textbf{Historical \& Process Information}
    \begin{itemize}
      \item Reasons for implementation choices
      \item Alternative approaches considered and rejected
      \item Change history beyond version control commits
      \item Stakeholder requirements that influenced the code
      \item Evolution of the code over time
      \item Original author's intent and mental model
    \end{itemize}
    
    \item \textbf{Performance \& Optimization Context}
    \begin{itemize}
      \item Performance characteristics and benchmarks
      \item Profiling data and bottlenecks
      \item Memory usage patterns
      \item Optimization opportunities
      \item Resource constraints
      \item Performance regressions over time
    \end{itemize}
    
    \item \textbf{Business \& Domain Knowledge}
    \begin{itemize}
      \item Business rules encoded in the algorithm
      \item Domain-specific terminology explanations
      \item Regulatory requirements implemented
      \item Business process mappings
      \item Customer use cases addressed
      \item Domain expert insights
    \end{itemize}
    
    \item \textbf{Testing \& Quality Context}
    \begin{itemize}
      \item Test coverage information
      \item Known edge cases
      \item Difficult-to-test scenarios
      \item Bug history and previous failures
      \item Validation approaches
      \item Quality metrics
    \end{itemize}
    
    \item \textbf{Runtime \& Environmental Context}
    \begin{itemize}
      \item Expected input/output examples
      \item Runtime behavior visualizations
      \item Environment dependencies
      \item Configuration requirements
      \item Deployment constraints
      \item Platform-specific considerations
    \end{itemize}
    
    \item \textbf{Security \& Compliance Information}
    \begin{itemize}
      \item Security considerations
      \item Privacy implications
      \item Compliance requirements
      \item Authentication/authorization context
      \item Data sensitivity information
      \item Security review history
    \end{itemize}
    
    \item \textbf{Maintenance \& Support Context}
    \begin{itemize}
      \item Known issues and workarounds
      \item Debugging tips
      \item Common maintenance tasks
      \item Support ticket references
      \item Troubleshooting guides
      \item Upgrade/migration considerations
    \end{itemize}
    
    \item \textbf{Educational \& Onboarding Context}
    \begin{itemize}
      \item Learning resources for new developers
      \item Step-by-step explanations
      \item Code walkthroughs
      \item Glossary of terms
      \item Interactive examples
      \item Conceptual models
    \end{itemize}
    
    \item \textbf{Collaborative Context}
    \begin{itemize}
      \item Code review comments and discussion
      \item Team member questions and answers
      \item Knowledge transfer notes
      \item Pair programming insights
      \item Cross-team coordination information
      \item Meeting notes relevant to the code
    \end{itemize}
    
    \item \textbf{Future Development Context}
    \begin{itemize}
      \item Planned enhancements
      \item Technical debt notes
      \item Roadmap items
      \item Future-proofing considerations
      \item Extensibility points
      \item Deprecation plans
    \end{itemize}
    
    \item \textbf{Visual Context}
    \begin{itemize}
      \item Data flow visualizations
      \item UI component renderings
      \item Color mappings and design assets
      \item Algorithm visualizations
      \item State machine diagrams
      \item Timeline visualizations
    \end{itemize}
    
    \item \textbf{External Relationships}
    \begin{itemize}
      \item Links to external documentation
      \item References to research papers or standards
      \item Connections to ticketing systems
      \item References to related codebases
      \item External API documentation
      \item Third-party library usage context
    \end{itemize}
  \end{itemize}
  
  \item \textbf{Examples in Real-World Software Development}
  \begin{itemize}
    \item \textbf{Comments \& Documentation}
    \begin{itemize}
      \item Code comments explaining complex logic or edge cases
      \item Function and class documentation
      \item README files and getting started guides
      \item API documentation and usage examples
    \end{itemize}
    
    \item \textbf{"Outside the Code" References}
    \begin{itemize}
      \item Stack Overflow answers bookmarked or copied
      \item Google/web search results you relied on
      \item Blog posts that helped solve a problem
      \item YouTube tutorials or conference talks
      \item Personal notes in text files or notebooks
    \end{itemize}
    
    \item \textbf{Visual Thinking}
    \begin{itemize}
      \item Whiteboard sketches you took photos of
      \item Diagrams drawn on paper or in tools like Miro
      \item Screenshots of working (or broken) features
      \item Napkin drawings that never made it into docs
      \item UI mockups and wireframes
    \end{itemize}
    
    \item \textbf{Knowledge from Others}
    \begin{itemize}
      \item Slack/Discord/Teams conversations with teammates
      \item Email threads explaining design decisions
      \item Code review comments that explained rationales
      \item Tribal knowledge shared verbally in meetings
      \item Pairing session insights
    \end{itemize}
    
    \item \textbf{Project Management Artifacts}
    \begin{itemize}
      \item Jira/Trello ticket details and requirements
      \item Pull request descriptions and discussions
      \item Sprint planning documents
      \item User stories and acceptance criteria
      \item Bug reports and their symptoms
    \end{itemize}
    
    \item \textbf{The "Why" Behind Code}
    \begin{itemize}
      \item Reasons for weird-looking code ("this looks strange but...")
      \item Business requirements that dictated unusual logic
      \item Workarounds for bugs in dependencies
      \item Performance hacks and their rationale
      \item Security considerations that affected design
    \end{itemize}
    
    \item \textbf{Testing Context}
    \begin{itemize}
      \item Test cases that drove implementation choices
      \item Edge cases discovered during testing
      \item Difficult-to-reproduce bugs
      \item User feedback that influenced the code
      \item Performance testing results
    \end{itemize}
    
    \item \textbf{"Remember This" Information}
    \begin{itemize}
      \item TODOs and FIXMEs
      \item Gotchas and pitfalls to watch out for
      \item Areas known to be fragile or in need of refactoring
      \item Reminders about deployment considerations
      \item Configuration requirements
    \end{itemize}
    
    \item \textbf{Environment \& Runtime Details}
    \begin{itemize}
      \item Local setup instructions
      \item Environment variables needed
      \item Database schema and relationships
      \item API endpoints and auth requirements
      \item Infrastructure dependencies
    \end{itemize}
    
    \item \textbf{Historical Context}
    \begin{itemize}
      \item "Why did we build it this way?"
      \item Previous approaches that failed
      \item Migration paths from legacy systems
      \item Compatibility requirements
      \item Technical debt explanations
    \end{itemize}
    
    \item \textbf{Future Plans}
    \begin{itemize}
      \item Planned refactoring ideas
      \item Feature roadmap items
      \item Performance improvement possibilities
      \item Scaling considerations for the future
      \item Ideas for better approaches
    \end{itemize}
    
    \item \textbf{Examples \& Sample Data}
    \begin{itemize}
      \item Sample inputs and expected outputs
      \item Test data sets
      \item Example configurations
      \item Demo scenarios
      \item User journey examples
    \end{itemize}
  \end{itemize}
\end{itemize}

\section{Demonstration Tasks}\label{sec:demo-examples}
During the round 2 demo, participants were shown the following material and asked to complete annotation tasks:
\begin{enumerate}
    \item An HTML file pre-annotated with a basic comment annotation type and another type that rendered a preview of a section of the HTML. All round 2 participants quickly confirmed at this stage that they understood what the annotations meant and how they worked.
    \item Another HTML file, where all round 2 participants quickly showed that they could add an annotation to the file without interviewer help.
    \item A JavaScript quicksort algorithm annotated with an animated GIF visualizing quicksort; no task was assigned for this example.
    \item A JavaScript regular expression (shown in Figure \ref{fig:teaser} on the left); participants were asked to add and use a Show Debugged Example annotation (see Section \ref{sec:program-data-live-execution}) to validate the regex. The extension's on- and offline annotation position updating capabilities were also demonstrated at this point.
    \item A Python program including an expression with floating-point error. Participants were asked to analyze the error and find a better expression using an \textit{Analyze Floating Point Expression} annotation type (Section \ref{sec:floating-point-analyzer}).
    \item A JavaScript loop body with an incorrect comment; participants were asked to add a unit test for the comment (see Section \ref{sec:lm-unit-test}).
\end{enumerate}

\section{Post-demo Surveys}\label{sec:post-survey}

Participants were asked the following questions in our post-demo survey. The questions mainly served as a starting point for discussion, and most participants skipped one or more questions due to time constraints.

\subsection{Round 1}
Question: How do the features currently provided by this system compare to your current tools for code documentation?	Are there particular projects you have worked on where you would like to have used the features of a tool like Codetations?

Question: What features and guarantees would the system have to provide before you would recommend the system to a team you were working with?

Question: [Discuss only] In the future, how would you expect this system to integrate with: 
\begin{itemize}
    \item AI code generation
    \item Version control
    \item Code review processes
    \item Team collaboration
    \item Continuous integration
    \item Other
\end{itemize}

Question: Orphaned annotations are annotations that no longer fit with the current document. In your opinion, how should the system handle re-attaching annotations that don't quite fit with the new document? Consider domains you have worked in and the types of changes to documents.

Question: For each annotation type you saw, was it useful as is or would it need more features to be useful? How difficult was it to understand the use of that annotation type?

Question: Which annotation types would you use in your daily work? Why?

Question: Can you imagine additional annotation types? Would you use any of these in your own work?

Question: [comment briefly for each] How do you imagine using a tool like Codetations as someone working in the following roles:
\begin{itemize}
    \item Original code author
    \item Code maintainer
    \item Tester
    \item Product manager
    \item API/library user
    \item Junior programmer
    \item Other roles you have worked in
\end{itemize}

Question: How easy would you say it was to learn to use Codetations to annotate code? (1-7)
Question: Discuss your idea of the "future of programming". Can you think of ways that annotations and annotation systems might have a role in this vision?

\subsection{Round 2}
Question: In what ways does this system address pain points in your current processes for working with contextual information? If needed, you can refer to the document on types of contextual information here.

Question: In what ways does this system NOT address pain points in your current processes for working with contextual information?

Question: How do the features of Codetations compare to your current documentation tools? In particular, compare Codetations to typical code comments.

Question: Which annotation types seemed most useful for workflows you are familiar with? 

Question: Are there specific projects or tasks where you would use a tool like Codetations?

Question: After seeing the demo, can you identify a recent situation where contextual information would have improved your work, and explain how this tool might host that information?

Question: What features would the system need before you'd recommend it to your team or use it personally? Are there any basic reasons you would hesitate to use it?

Question: On a scale of 1-7, if a polished version were available through your IDE, how likely would you be to use this tool?

\section{Other Applications}\label{sec:other-applications}
\subsection{Recording Structured LLM Interactions}

LLMs have increasingly become tools of choice for low-code programming, aiding developers of various skill levels in the rapid development of software by generating code snippets from high-level descriptions. However, observations by Liu et al. \cite{liu2024empirical} highlight user concerns that LLMs may introduce variability in the correctness and quality of the generated code. Bird et al. further note the establishment of code provenance as a key issue in systems that use generated code \cite{bird2022taking}, i.e., when code is generated in a chat interface, the conversation that led to the code, including valuable specification data, is typically not attached to the codebase with the resulting code.

We could easily foresee simply recording unstructured chats with an LLM as annotations, already improving upon current practices. This workflow could be extended to any other tool that generates or validates code. For example, the user's work in Ferdowsi et al.'s LEAP system \cite{ferdowsi2023live}, which lets users validate AI-generated code with live feedback on the line-level behavior of generated code, could be preserved in a similar way.

\section{More Worked Examples}\label{sec:worked-examples}

Here we include other worked examples exploring the effects of additional context on an LLM's program repair performance for debugging problems generated by the model itself.

\subsection{Real-World Assumptions} \label{ssec:realworld}
Many programming errors stem not from faulty logic within the code itself, but from real-world constraints that lie outside the immediate scope of the code snippet. These constraints --- whether hardware-specific quirks, environmental dependencies, or undocumented design choices --- are often tacitly understood by human teams but remain opaque to LLMs. Just as human programmers rely on experience and tribal knowledge to navigate such issues, LLMs require similar cues to surface relevant latent knowledge.

Consider the NXP PN532 NFC chip, widely used via Adafruit breakout boards. When configured in UART Normal Mode, the chip expects an ACK byte from the host after every frame. By default, the Adafruit library disables this handshake, leading to intermittent data loss; this is a subtle bug that eludes detection unless agents are familiar with the underlying hardware protocol. A correct fix involves switching the chip into High-Speed UART (HSU) mode, which bypasses the need for ACKs entirely.

When prompted with a three-line NFC reading function and a vague error report describing occasionally dropped data, GPT-4o responded by dramatically expanding the code: introducing delays, buffer resets, and packet-length guards. Despite these elaborate modifications, the bug remained. However, when the same code was presented alongside data from a Codetations Show Debugged Example tool demonstrating the failure and paired with distractor documentation about the PN532 chip, the model rapidly identified the true issue and recommended the correct configuration change.

\end{document}